\documentclass[12pt]{article}
\usepackage{cite}
\usepackage{graphicx}

\input epsf.sty
\usepackage{pinlabel}

\usepackage{graphicx}

\usepackage{latexsym}

 \usepackage{amsmath}
 \usepackage{amssymb}
 \usepackage{amsthm}

\newcommand{\pic}[2]{\raisebox{-.4\height}{\includegraphics[scale=#2]{#1}}}

\def\Xor{\pic{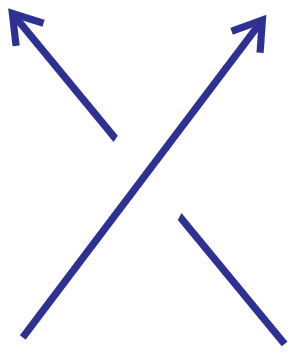}{.300}}
\def\Yor{\pic{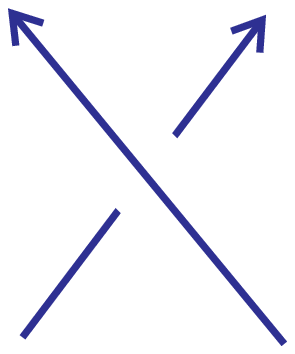} {.300}}
\def\Ior{\pic{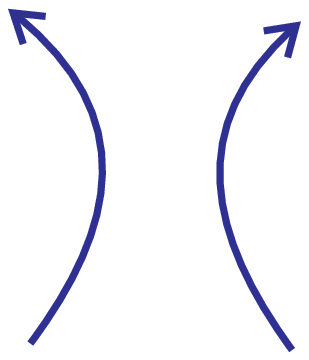} {.300}}

\def\Rcurlor{\pic{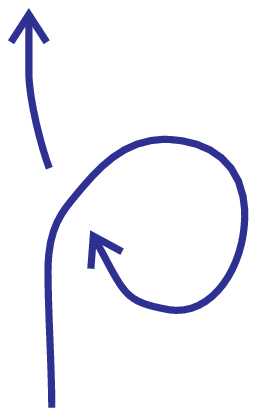}{.300}}
\def\Lcurlor{\pic{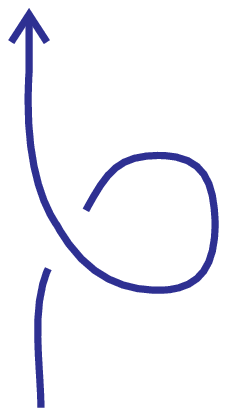} {.300}}
\def\Idor{\pic{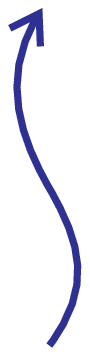} {.300}}
\def\unknot{\pic{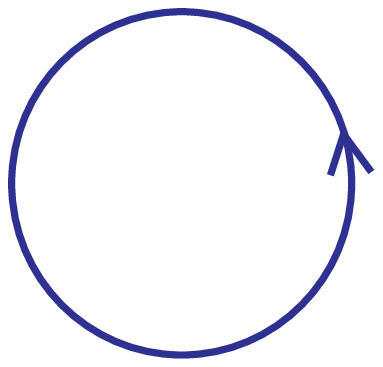} {.200}}
\def\ununknot{\pic{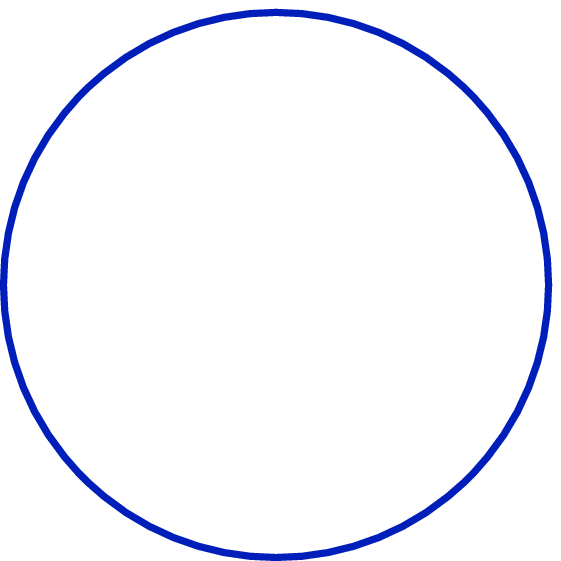} {.135}}

\def\Xunor{\pic{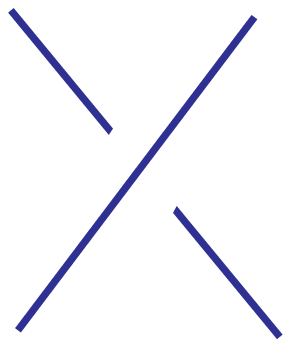}{.300}}
\def\Yunor{\pic{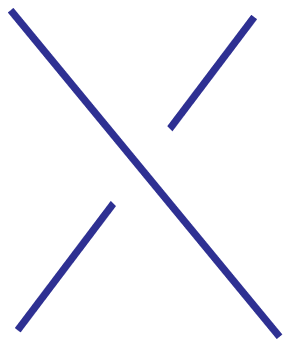} {.300}}
\def\Iunor{\pic{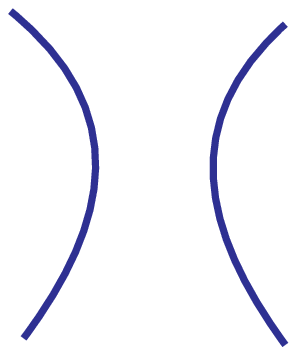} {.300}}
\def\Infunor{\pic{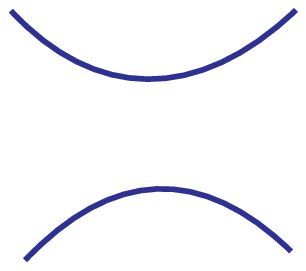} {.300}}

\def\Rcurlunor{\pic{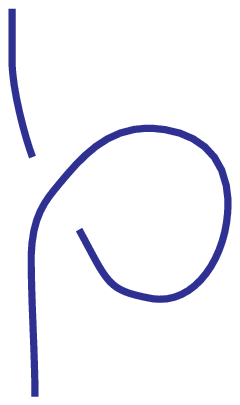}{.300}}
\def\Lcurlunor{\pic{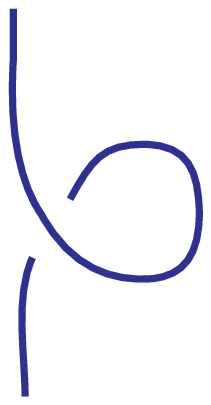} {.300}}
\def\Idunor{\pic{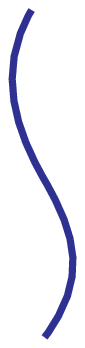} {.300}}

\newtheorem{theorem}{Theorem}[section]

\theoremstyle{definition}

\newtheorem{example}[theorem]{Example}

\newtheorem{conjecture}{Conjecture}[section]

\newtheorem{remark}[theorem]{Remark}

\newcommand{\CB}{{\cal B}}

\newcommand{\CG}{{\cal G}}
\newcommand{\CH}{{\cal H}}
\newcommand{\CI}{{\cal I}}

\newcommand{\CK}{{\cal K}}
\newcommand{\CL}{{\cal L}}

\newcommand{\CN}{{\cal N}}
\newcommand{\CO}{{\cal O}}

\newcommand{\CR}{{\cal R}}

\def\IZ{{\mathbb Z}}
\def\IR{{\mathbb R}}
\def\IC{{\mathbb C}}
\def\IP{{\mathbb P}}

\def\IS{{\mathbb S}}
\def\IQ{{\mathbb Q}}

\newcommand{\tr}{{\rm Tr}}

\newcommand{\be}{\begin{equation}}
\newcommand{\ee}{\end{equation}}
\newcommand{\ba}{\begin{aligned}}
\newcommand{\ea}{\end{aligned}}
\newcommand{\ben}{\begin{eqnarray}\displaystyle}
\newcommand{\een}{\end{eqnarray}}

\newcommand{\sectiono}[1]{\section{#1}\setcounter{equation}{0}}

\newdimen\tableauside\tableauside=1.0ex
\newdimen\tableaurule\tableaurule=0.4pt
\newdimen\tableaustep
\def\phantomhrule#1{\hbox{\vbox to0pt{\hrule height\tableaurule width#1\vss}}}
\def\phantomvrule#1{\vbox{\hbox to0pt{\vrule width\tableaurule height#1\hss}}}
\def\sqr{\vbox{%
  \phantomhrule\tableaustep
  \hbox{\phantomvrule\tableaustep\kern\tableaustep\phantomvrule\tableaustep}%
  \hbox{\vbox{\phantomhrule\tableauside}\kern-\tableaurule}}}
\def\squares#1{\hbox{\count0=#1\noindent\loop\sqr
  \advance\count0 by-1 \ifnum\count0>0\repeat}}
\def\tableau#1{\vcenter{\offinterlineskip
  \tableaustep=\tableauside\advance\tableaustep by-\tableaurule
  \kern\normallineskip\hbox
    {\kern\normallineskip\vbox
      {\gettableau#1 0 }%
     \kern\normallineskip\kern\tableaurule}%
  \kern\normallineskip\kern\tableaurule}}
\def\gettableau#1{\ifnum#1=0\let\next=\null\else
\squares{#1}\let\next=\gettableau\fi\next}

\newcommand{\figref}[1]{Fig.~\protect\ref{#1}}

\begin{document}

\begin{titlepage}

\vskip 3cm

\centerline{\Large \bf
String theory and the Kauffman polynomial}

\medskip

\vspace*{4.0ex}

\centerline{\large \rm
Marcos Mari\~no }

\vspace*{4.0ex}

\centerline{ D\'epartement de Physique Th\'eorique et Section de Math\'ematiques}
\centerline{ Universit\'e de Gen\`eve}
\centerline{  Gen\`eve, CH-1211 Switzerland}
\vspace*{2.0ex}
\centerline{marcos.marino@unige.ch}

\vspace*{10.0ex}

\centerline{\bf Abstract} \bigskip

We propose a new, precise integrality conjecture for the colored Kauffman polynomial of knots and links inspired by large $N$ dualities and the structure of 
topological string theory on 
orientifolds. According to this conjecture, the natural knot invariant in an unoriented theory involves both the colored Kauffman polynomial 
and the colored HOMFLY polynomial for composite representations, i.e. it involves the full 
HOMFLY skein of the annulus. The conjecture sheds new light on the relationship between the Kauffman and the HOMFLY polynomials, and it implies 
for example Rudolph's theorem. We provide various non-trivial tests of the conjecture and we sketch the string theory arguments that lead to it. 

\vfill \eject

\end{titlepage}

\tableofcontents

\sectiono{Introduction}

The HOMFLY \cite{homfly} and the Kauffman \cite{kauffman} 
polynomials are probably the most useful two-variable polynomial invariants of knots and links. 
Both of them generalize the Jones polynomial, and they have become basic building blocks of 
quantum topology. However, many aspects of these polynomial invariants are still poorly understood. As Joan Birman 
remarked in 1993, ``we can compute the simplest of the invariants by hand and quickly fill pages (...) without 
having the slightest idea what they mean" \cite{birman}. 

One particular interesting question concerns 
the relationship between the HOMFLY and the Kauffman invariants. Since their discovery almost thirty years ago, a number of isolated connections have been found between them. 
For example, when written in an appropriate way, they have the same lowest order term \cite{lickorish}. Other connections can be found when one considers their {\it colored} versions. The colored invariants can be formulated in terms of skein theory, in terms of quantum groups, or in terms of Chern--Simons gauge theory \cite{wittenjones}. In the language of quantum groups or Chern--Simons theory, different colorings correspond to different choices of group representation. The original HOMFLY and Kauffman invariants are obtained when one considers the fundamental representation of ${\rm SU}(N)$ and ${\rm SO}(N)/{\rm Sp}(N)$, respectively. One could consider other 
representations, like for example the adjoint representation. An intriguing result of Rudolph \cite{rudolph} states that the HOMFLY invariant of a link colored by the adjoint representation equals the square of the Kauffman polynomial of the same link, after the coefficients are reduced modulo two. This type of relationship has been recently extended by 
Morton and Ryder to more general colorings \cite{mortontangle, mr}. In spite of these connections, no unified, general picture has emerged to describe both invariants.

More recently, knot invariants have been reinterpreted in the context of string theory thanks to 
the Gopakumar--Vafa conjecture \cite{gv}, which postulates an equivalence between the $1/N$ expansion of 
Chern--Simons theory on the three-sphere, and topological string theory on a Calabi--Yau manifold called 
the resolved conifold. As a consequence of this conjecture, correlation functions of Chern--Simons gauge theory with ${\rm U}(N)$ gauge group (i.e. colored HOMFLY invariants) are given by correlation 
functions in open topological string theory, which mathematically correspond to open Gromov--Witten invariants. Since Gromov--Witten 
invariants enjoy highly nontrivial integrality properties \cite{gvonetwo,ov,lmv}, this equivalence provides strong structural results on the coloured HOMFLY polynomial \cite{ov,lmv,lm} which have been 
tested in detail in various cases \cite{lm,rs,lz} and finally proved in \cite{lpeng}. 
Moreover, there is a full cohomology theory behind these invariants \cite{lmv} which should be 
connected to categorifications of the HOMFLY polynomial \cite{gsv}. Therefore, the string theory description ``explains" to a large extent many aspects of the colored ${\rm U}(N)$ invariants and leads to new predictions about their algebraic structure. 

The string theory perspective is potentially the most powerful tool to understand the connections between the colored HOMFLY and Kauffman polynomials. From the point of view of 
Chern--Simons theory, these polynomials correspond to the gauge groups ${\rm U}(N)$ and ${\rm SO}(N)/{\rm Sp}(N)$, respectively. 
But when a gauge theory has a string theory large $N$ dual, as it is the case here, the theory with orthogonal or symplectic gauge groups can be obtained from the theory with unitary gauge group by using a special type of orbifold action called an {\it orientifold}. The building block of an orientifold is an involution $\CI$ in the target space $X$ of the string theory, which is then combined with an orientation reversal in the worldsheet of the string to produce unoriented strings in the quotient space $X/\CI$. Very roughly, one finds that correlation functions in the orbifold theory are given by correlation functions in the ${\rm SO}/{\rm Sp}$ gauge theories. As for any orbifold, these functions are given by a sum over an ``untwisted" sector involving oriented strings, and a ``twisted" sector involving the unoriented strings introduced by the orientifold. The contributions from oriented strings are still given by correlation functions in the ${\rm U}(N)$ gauge theory. 

The use of orientifolds in the context of the Gopakumar--Vafa conjecture was initiated in 
\cite{sv}, which identified the relevant involution of the resolved conifold and 
studied the closed string sector. This line of research was developed in more detail in \cite{bfmone,bfmtwo}. In particular, \cite{bfmtwo} extended the orientifold action to the open string sector and pointed out that, as a consequence of the underlying string/gauge theory correspondence, the colored Kauffman invariant of a link should be given by the sum of an appropriate HOMFLY invariant plus an ``unoriented" contribution. The results of \cite{bfmtwo} made possible to formulate some partial conjectures on the structure of the Kauffman polynomial and test them in examples (see \cite{br} for further tests)\footnote{Some of the proposals of \cite{bfmtwo} were reformulated and recently proved in \cite{rch}.}. Unfortunately, these results were not precise enough to provide a full, detailed string-based picture. The reason was that one of the crucial ingredients --the appropriate 
HOMFLY invariant that corresponds to the untwisted sector of the orientifold-- was not identified. 

In this paper we remedy this situation and we identify these invariants as HOMFLY polynomials colored by {\it composite} representations 
of ${\rm U}(N)$. This will allow us to state a precise conjecture on the structure of the coloured Kauffman polynomial of knots and links. In skein-theoretic language, the appearance of composite representations means that, in order to understand the colored Kauffman polynomial in the light of string theory, one has to use the full HOMFLY skein of the annulus (see for example \cite{hm}). We will indeed see that 
the natural link invariant to consider in an unoriented theory involves both the colored Kauffman polynomial 
and the colored HOMFLY polynomial for composite representations and for all possible orientations of the link components. Our conjecture generalizes the results of \cite{ov,lmv} for the ${\rm U}(N)$ case, and it ``explains" various aspects of the relationship between the HOMFLY and the Kauffman polynomials, like 
for example Rudolph's theorem. It also predicts some new, simple relationships between the Kauffman and the HOMFLY polynomial of links. 

In terms of open topological string theory, this paper adds little to the results of \cite{bfmtwo}. The bulk of the paper is then 
devoted to a detailed statement and discussion of the conjecture in the language of knot theory. Section 2 introduces our notation and 
reviews the construction of the colored HOMFLY and Kauffman polynomials, as well as 
of their relations. In section 3 we review the conjecture of \cite{ov,lmv} and we state the new conjecture 
for the colored Kauffman polynomial. Section 4 provides some nontrivial evidence for the conjecture 
by looking at particular knots and links, and it explains how some standard results relating the HOMFLY and the 
Kauffman polynomials follow easily from our conjecture. 
In section 5 we sketch the string theory arguments that lead to the conjecture, 
building on \cite{sv,bfmone,bfmtwo}. Finally, section 6 contains some conclusions and prospects for future work.

\sectiono{Colored HOMFLY and Kauffman polynomials}

In this section we introduce various tools from the theory of symmetric polynomials and we recall 
the construction of the colored Kauffman and HOMFLY polynomials, mainly to fix notations. 

\subsection{Basic ingredients from representation theory}

Let $R$ be an irreducible representation of the symmetric group $S_{\ell}$. We will represent it by a Young diagram or partition,
\be
R=\{ l_i\}_{i=1,\cdots, r(R)}, \qquad l_1 \ge l_2 \ge \cdots l_{r(R)}
\ee
where $l_i$ is the number of boxes in the $i$-th row of the diagram and $r(R)$ is the total number of rows. Important quantities associated to the 
diagram are its total number of boxes, 
\be
\ell(R) =\sum_{i=1}^{r(R)} l_i
\ee
which equals $\ell$, as well as the quantity
\be
\kappa_R =\sum_{i=1}^{r(R)} l_i(l_i-2i+1).
\ee

The ring of symmetric polynomials in an infinite number of variables $\{ v_i\}_{i\ge 1}$ will be denoted by $\Lambda$. It can be easily constructed as a direct limit of the ring 
of symmetric polynomials with a finite number of variables, see for example \cite{macdonald} for the details. 
It has a basis given by the Schur polynomials $s_R (v)$, which are labelled by Young diagrams. The multiplication rule for these polynomials is encoded in the 
Littlewood--Richardson coefficients
\be
\label{lr}
s_{R_1} (v) s_{R_2} (v) =\sum_R N_{R_1 R_2}^R s_R (v). 
\ee
The identity of this ring is the Schur polynomial associated to the empty diagram, which we will denote by $R=\cdot$. 
We will also need the $n$-th Adams operation 
\be
 \psi_n(s_R(v)) = s_R(v^n)=s_R(v_1^n, v_2^n, \cdots)
\ee
One can use elementary representation theory of the symmetric group to express $s_R(v^n)$ as a linear combination of Schur polynomials 
labelled by representations $U$ with $n\cdot \ell(R)$ boxes
\be
s_R(v^n) =\sum_U c_{n;R}^U \, s_U(v).
\ee
Let $\chi_R$ be the character of the symmetric group associated to the diagram $R$. Let $C_{\mu}$ be the conjugacy class 
associated to the partition $\mu$, and let $|C_{\mu}|$ be the number of elements in the conjugacy class. It is easy to show that the coefficients $c_{n;R}^U$ are given by \cite{lz}
\be
\label{cnr}
c_{n;R}^U =\sum_\mu {1\over z_{\mu}} \chi_R(C_{\mu})\chi_U(C_{n\mu})
\ee
where $n\mu=(n\mu_1, n\mu_2,\cdots)$ and 
\be
z_\mu ={\ell(\mu)! \over |C_{\mu}|}.
\ee
We will also regard a Young diagram $R$ as an irreducible representation of ${\rm U}(N)$. The quadratic Casimir of $R$ is then given by 
\be
C_R = \kappa_R + N \ell(R).
\ee

\begin{figure}[!ht]
\leavevmode
\begin{center}
\epsfysize=4cm
\epsfbox{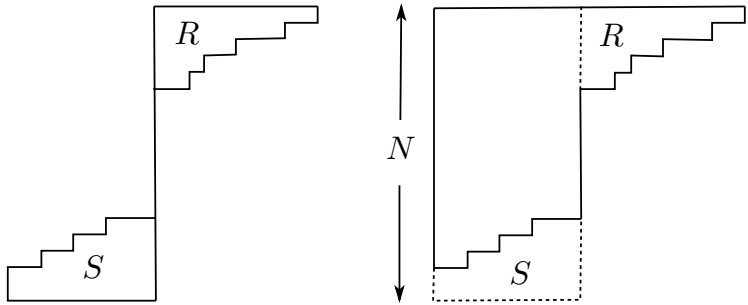}
\end{center}
\caption{A composite representation made out of the diagrams $R$ and $S$. }
\label{compositefig}
\end{figure} 

The most general irreducible representation of ${\rm U}(N)$ is a {\it composite} representation (see for example \cite{jap} for a collection of useful results on composite representations). 
Composite representations are labelled by a pair of Young diagrams
\be
\label{composte}
(R, S).
\ee
This representation is usually depicted as in the left hand side of \figref{compositefig}, where the second representation $S$ is drawn upside down at the 
bottom of the diagram. When regarded as 
a representation of ${\rm SU}(N)$, the composite representation corresponds to the diagram depicted on the right hand side of \figref{compositefig}, and it has in total
\be
N \mu_1 +\ell(R) -\ell(S)
\ee
boxes, where $\mu_1$ is the number of boxes in the first row of $S$. For example, the composite representation $(\tableau{1}, \tableau{1})$ is the adjoint representation of 
${\rm SU}(N)$. It is easy to show that \cite{gt}
\be
C_{(R, S)} =C_{R} +C_{S}. 
\ee
The composite representation can be understood as the tensor product $R\otimes \overline S$, where $\overline S$ is the conjugate representation to $S$, plus a series of ``lower order corrections" involving tensor products of smaller representations. The precise formula 
is \cite{jap}
\be
\label{tensorcomposite}
(R,S)=\sum_{ U, V,W} (-1)^{\ell(U)} N^R_{U V} N^S_{U^T W}\, (V\otimes {\overline W}).
\ee

\subsection{The colored HOMFLY polynomial} 

The HOMFLY polynomial of an oriented link $\CL$, $P_{\CL}(t, \nu)$, can be defined by using a planar projection of $\CL$. This gives an oriented diagram in the plane which will be denoted 
as $D_{\CL}$. The skein of the plane is the set of linear combinations of these diagrams, modulo the skein relations
\be
\ba
\Xor &-\Yor  \quad=\quad (t-t^{-1}) \Ior,\\
\Rcurlor\quad &= \quad \nu \ \Idor\ , \qquad \Lcurlor\quad=\quad \nu^{-1}\  \Idor 
\ea
\ee
Using the skein relations, the diagram of a link $D_{\CL}$ can be seen to be proportional to the trivial diagram \unknot. The proportionality factor $\langle D_{\CL} \rangle $ is a scalar and gives a 
regular isotopy invariant (i.e. a quantity which is invariant under the Reidemeister moves II and III, but not under the I). A true ambient isotopy invariant is obtained by defining
\be
\label{homflydef}
P_{\CL}(t, \nu) = \nu^{-\overline w(D_{\CL})} \langle D_{\CL} \rangle 
\ee
where $\overline w(D_{\CL})$ is the self-writhe of the link diagram $D$ (see for example \cite{lickorish}, p. 173). This is defined as the sum of the signs of crossings at which all 
link components cross themselves, and not other components. It differs from the standard writhe $w(D)$ in twice the total linking number of the 
link, ${\rm lk}(\CL)$. Notice that the standard HOMFLY polynomial is usually defined by using the total writhe in (\ref{homflydef}) \cite{lickorish}. Therefore, the HOMFLY polynomial, as defined in (\ref{homflydef}), is given by the standard HOMFLY polynomial times a factor 
\be
\label{nulinking}
\nu^{2 \, {\rm lk}(\CL)}. 
\ee
As discussed in detail in \cite{lmv}, this is the natural version of the HOMFLY polynomial from the string theory point of view. 
The HOMFLY invariant of the link is then defined by 
\be
\label{hinv}
\CH (\CL) =P_{\CL}(t, \nu) \CH \Bigl( \unknot\Bigr), 
\ee
and we choose the normalization 
\be
\CH\Bigl( \unknot\Bigr) ={\nu-\nu^{-1} \over t-t^{-1}}. 
\ee

\begin{figure}[!ht]
\leavevmode
\begin{center}
\epsfysize=7cm
\epsfbox{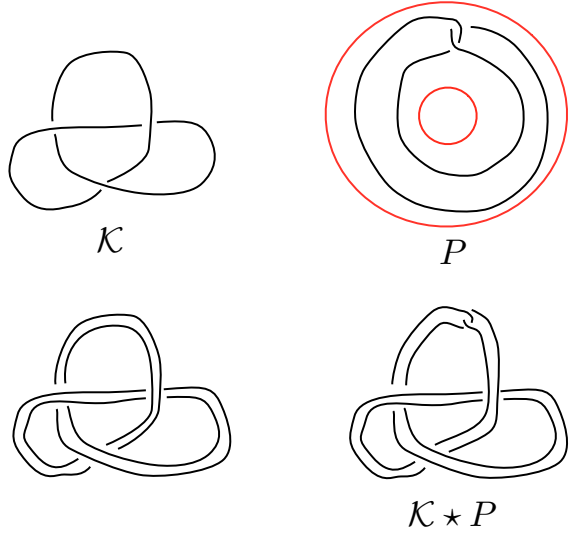}
\end{center}
\caption{An example of a satellite knot. The companion knot $\CK$ is the trefoil knot, and below we show the framing annulus. If we replace this framing annulus by the pattern $P$, 
we obtain the satellite $\CK \star P$. }
\label{satellite}
\end{figure}

From the skein theory point of view, the colored HOMFLY invariant of a link is obtained by considering {\it satellites} of the knot. 
Let $\CK$ be a framed knot, and let $P$ be a knot diagram in the annulus. Around $\CK$ there is a 
framing annulus, or equivalently a parallel of $\CK$. The {\it satellite knot}
\be
\CK\star P
\ee
is obtained by replacing the framing annulus around $\CK$ by $P$, or equivalently by mapping $P$ to $\IS^3$ using the parallel of $\CK$. Here, 
$\CK$ is called the {\it companion knot} while $P$ is called the {\it pattern}. In \figref{satellite} we show a satellite where the companion $\CK$ is the trefoil knot. 

Since the diagrams in the annulus form a vector space (called the skein of the annulus) we can obtain the most general satellite of a knot by considering 
the basis of this vector space. There is a very convenient basis constructed in \cite{hm} whose elements are labelled by pairs of Young diagrams 
$P_{(R, S)}$. Given a knot $\CK$, 
the HOMFLY invariant colored by the partitions $(R, S)$ is simply 
\be
\CH_{(R, S)} (\CK) = \CH \bigl(\CK\star P_{(R, S)}\bigr).
\ee
If we have a link $\CL$ with $L$ components $\CK_1, \cdots, \CK_L$, one can color each component independently, and one obtains an invariant of the form 
\be
\label{colHlink}
\CH_{(R_1, S_1), \cdots, (R_L, S_L)}(\CL).
\ee

\begin{figure}[!ht]
\leavevmode
\begin{center}
\epsfysize=3.5cm
\epsfbox{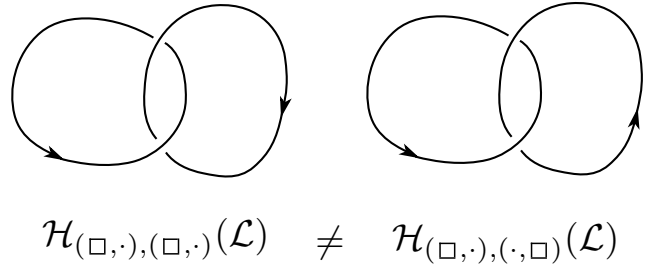}
\end{center}
\caption{Changing $(R,S)$ into $(S,R)$ reverses the orientation of a knot. If the knot is a component of a link, this leads in general to different HOMFLY invariants.}
\label{hreversed}
\end{figure} 

The colored HOMFLY invariant has various important properties which will be needed in the following:
\begin{enumerate}

\item The pattern $P_{(S,R)}$ is equal to the pattern $P_{(R, S)}$ with its orientation reversed. In particular, coloring with $(\tableau{1},\cdot)$ gives the original knot $\CK$, while coloring with $(\cdot, \tableau{1})$ gives the knot $\overline{\CK}$ with the {\it opposite} orientation. Since the HOMFLY invariant of a knot is invariant under reversal of orientation, we have that
\be
\label{hcomin}
\CH_{(S,R)}(\CK)=\CH_{(R,S)}(\overline{\CK})=\CH_{(R,S)}(\CK). 
\ee
However, the HOMFLY invariant of a link is only invariant under a {\it global} reversal of orientation, therefore in general one has that
\be
\label{reversal}
\CH_{(R_1, S_1), \cdots, (S_j, R_j), \cdots, (R_L, S_L)}(\CL) =\CH_{(R_1, S_1), \cdots, (R_j, S_j), \cdots, (R_L, S_L)}({\overline \CL}_j),
\ee
where ${\overline \CL}_j$ is the link obtained from the link $\CL$ by reversing the orientation of the $j$-th component, see for example \figref{hreversed}. 

\item If one of the patterns is empty, say $S=\cdot$, the skein theory is simpler and it has been developed in for example \cite{am,aiston}. In this case, the HOMFLY invariant of the knot 
$\CK$ (which we denote by $\CH_R (\CK)$) is equal to the invariant of $\CK$ obtained from the quantum 
group $U_q({\rm sl}(N,\IC))$ in the representation $R$, with the identification 
\be
t=q^{1/2}, \quad \nu =t^N. 
\ee
In particular we have
\be
\CH_R \Bigl(\unknot\Bigr)={\rm dim}_q\, R
\ee
where ${\rm dim}_q\, R$ is the quantum dimension of $R$. 

 \item For a general pattern labeled by two representations $(R,S)$, the HOMFLY invariant of the knot $\CK$, $\CH_{(R,S)}(\CK)$ equals the invariant of $\CK$ obtained from 
 the quantum group $U_q({\rm gl}(N,\IC))$ in the composite representation $(R,S)$. In particular \cite{hm}
 \be
 \label{qdimcr} 
 \CH_{(R,S)} \Bigl(\unknot\Bigr)={\rm dim}_q\, (R,S) = \sum_{ U, V,W} (-1)^{\ell(U)} N^R_{U V} N^S_{U^T W}\Bigl( {\rm dim}_q  V \Bigr) \Bigl( {\rm dim}_q  W \Bigr).
 \ee
 Here, $U^T$ is the transposed Young diagram. The second equality follows from (\ref{tensorcomposite}). 
 
 \end{enumerate}
 \begin{figure}[!ht]
\leavevmode
\begin{center}
\epsfysize=7cm
\epsfbox{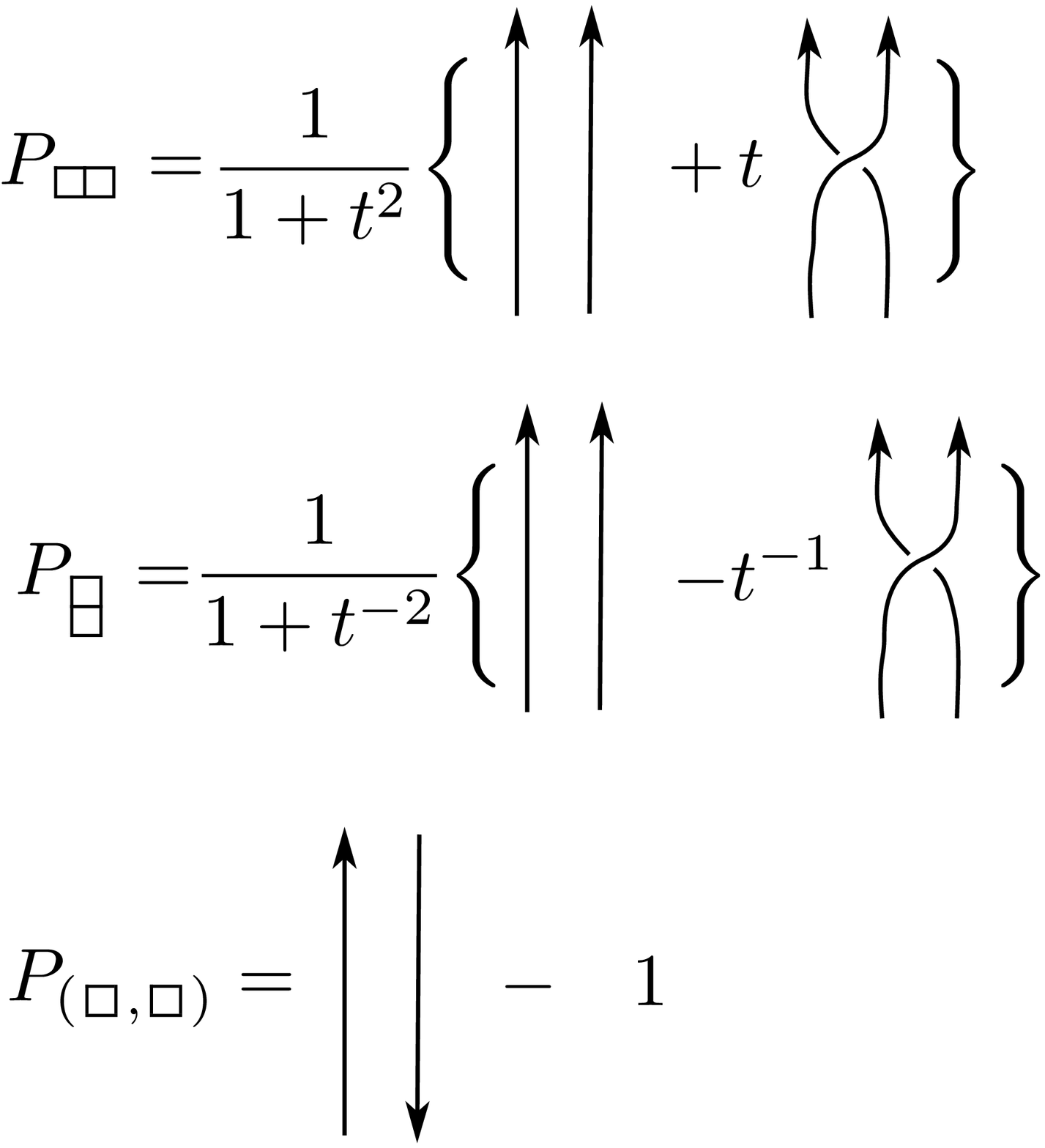}
\end{center}
\caption{Examples of patterns for various representations. The patterns are written as formal combinations of braids, and after closing them we find elements in the skein of the annulus. In the last 
example, $1$ refers to the empty diagram.}
\label{patterns}
\end{figure} 
In \figref{patterns} we show some examples of patterns associated to different representations. The patterns are represented as elements in the braid group, which can be closed by joining the endpoints to produce patterns in the annulus. 

In the following we will denote
\be
z=t-t^{-1}.
\ee
\begin{remark} In this paper, the above skein rules will be used to compute the 
values of the HOMFLY and Kauffman invariants in the {\it standard} framing. Starting from this framing, a change of framing by $f$ units is done 
through \cite{mv}
\be
\label{framing}
\CH_R(\CK) \rightarrow (-1)^{f \ell(R)} t^{f\kappa_R} \CH_R(\CK).
\ee
This is the rule that preserves the integrality properties of the invariants that will be discussed below. 
The framing of links is done in a similar way, with one framing factor like (\ref{framing}) for each component. 
\end{remark}

\begin{example} The HOMFLY polynomial of the trefoil knot is, in our conventions, 
\be
P_{3_1}(t,\nu)= 2\nu^2 -\nu^4 -z^2 \nu^2,
\ee
while the HOMFLY polynomial of the Hopf link is
\be
P_{2_1^2} (t,\nu)= \bigl( \nu-\nu^{-1} \bigr) z^{-1}+ \nu z.
\ee
\end{example}

\subsection{The colored Kauffman polynomial} 
The Kauffman polynomial is also defined by a skein theory \cite{kauffman}, but the diagrams correspond now to planar projections of {\it unoriented} knots and links. 
The skein relations are
\be
\ba
\Xunor\quad &-\quad\Yunor \quad=\quad (t-t^{-1}) \left(\, \Iunor-\Infunor\right),\\
\Rcurlunor\quad &= \quad \nu \ \Idunor\ , \qquad \Lcurlunor\quad=\quad \nu^{-1}\  \Idunor\  
\ea
\ee
This is sometimes called the ``Dubrovnik" version of the Kauffman invariant. As in the case of HOMFLY, the diagram of an unoriented link $\CL$, which will be denoted by $E_{\CL}$, is proportional to the trivial diagram \ununknot, and the proportionality factor $\langle E_{\CL}\rangle$ is a regular 
isotopy invariant. The Kauffman polynomial is defined as
\be
F_{\CL}(t,\nu)=\nu^{-\overline w(E_{\CL})}\langle E_{\CL}\rangle.
\ee
Like before, this differs from the standard Kauffman polynomial (as defined for example in \cite{lickorish}) in an overall factor (\ref{nulinking}). 
More importantly, the use of the self-writhe guarantees that the resulting polynomial is still an invariant of unoriented links. 
The Kauffman invariant of the link will be defined as
\be
\label{kinv}
\CG (\CL) = F_{\CL}(t,\nu) \CG \Bigl(\ununknot \Bigr), 
\ee
and we choose the normalization 
\be
\CG \Bigl(\ununknot \Bigr) =1+{\nu -\nu^{-1} \over t-t^{-1}}.
\ee
The colored Kauffman polynomial is obtained, similarly to the HOMFLY case, by considering the Kauffman skein of the annulus and by forming satellites with elements of this skein taken as patterns. 
There is again a basis $y_R$ labelled by Young tableaux \cite{bb}, and we define
\be
\CG_R (\CK)=\CG(\CK\star y_R). 
\ee
For a link of $L$ components, we can color each component independently, and one obtains in this way the colored Kauffman invariant of the link
\be
\CG_{R_1, \cdots, R_L} (t, \nu). 
\ee
The invariant defined in this way equals the invariant obtained from the quantum group $U_q({\rm so}(N, \IC))$ in the representation $R$, after identifying 
\be
t=q^{1/2}, \qquad \nu=q^{N-1}.
\ee
In particular, for the unknot the invariant is equal to the quantum dimension of $R$,
\be
\CG_R\Bigl( \ununknot \Bigr) ={\rm dim}^{\rm SO(N)}_q \, R.
\ee
The results for the Kauffman invariants of knots and links will be presented in the standard framing. The change of framing is also done with the rule (\ref{framing}). 

\begin{example} The Kauffman polynomial of the trefoil knot is, in our conventions, 
\be
F_{3_1}(t,\nu)=2\nu^2 -\nu^4 +z(-\nu^3 +\nu^5) + z^2 (\nu^2-\nu^4),
\ee
while that of the Hopf link is
\be
\label{khopf}
F_{2_1^2}(t,\nu)=z^{-1} \Bigl\{ \nu -\nu^{-1} + z +  z^2 (\nu -\nu^{-1})\Bigr\}.
\ee
\end{example}

\subsection{Relationships between the HOMFLY and the Kauffman invariants}
As we mentioned in the introduction, the colored HOMFLY and Kauffman invariants of a link are not unrelated. 
The simplest relation concerns the invariants of a link $\CL$ in which all components have the coloring $R=\tableau{1}$, i.e. the original HOMFLY and Kauffman polynomials. It is easy to show that these polynomials have the structure
\be
\label{structure}
P_{\CL} (z, \nu)=z^{1-L} \sum_{i\ge 0} p_i(\nu) z^{2i}, \qquad F_{\CL} (z,\nu)=z^{1-L} \sum_{i\ge0} k_i(\nu) z^i.
\ee
It turns out that (see for example \cite{lickorish}, Proposition 16.9)
\be
\label{pk}
p_{0}(\nu)=k_{0} (\nu).
\ee
In general, the Kauffman polynomial contains many more terms than the HOMFLY polynomial. 
In particular, as (\ref{structure}) shows, it contains both even and odd powers of $z$, while the HOMFLY 
polynomial only contains even powers. In the case of torus knots the HOMFLY polynomial can even be obtained from the 
Kauffman polynomial by the formula \cite{lp}
\be
\label{lprelation}
\CH(z, \nu)={1\over 2} (\CG (z,\nu) -\CG(-z,\nu)).
\ee

There are also highly nontrivial relations between the two invariants when we consider colorings. An intriguing theorem of Rudolph \cite{rudolph} states the following. Let $\CL$ be an unoriented link with $L$ components. Pick an arbitrary orientation of $\CL$ and 
consider its HOMFLY invariant
\be
\label{unorHinv}
\CH_{(\tableau{1}, \tableau{1}), \cdots, (\tableau{1}, \tableau{1})}(\CL). 
\ee
Due to (\ref{reversal}), this invariant does not depend on the choice of orientation in $\CL$, and it is therefore an invariant of the unoriented link. 
One can show (\ref{unorHinv}) is an element in $\IZ[z^{\pm 1}$, $\nu^{\pm1}]$ (see for example \cite{mortontangle}). The square of the Kauffman invariant of $\CL$, $\CG^2(\CL)$, belongs to the same ring. By reducing the coefficients of these polynomial modulo $2$, we obtain two polynomials in $\IZ_2[z^{\pm1} ,\nu^{\pm 1}]$. Rudolph's theorem states that these reduced polynomials are the same. In other words, 
\be
\label{rudolphth}
\CG^2(\CL) \equiv\CH_{(\tableau{1}, \tableau{1}), \cdots, (\tableau{1}, \tableau{1})} (\CL) \quad {\rm mod}\, \ 2,
\ee
see \cite{ryder} for this statement of Rudolph's theorem. Morton and Ryder have recently extended this result to more general colorings \cite{mortontangle, mr}. This generalization requires more care since now the invariants have denominators involving products of $t^r-t^{-r}$, $r\in \IZ_{>0}$. However, one can still make sense of the reduction modulo $2$, and 
one obtains that, for any unoriented link $\CL$, 
\be
\label{mrgen}
\CG^2_{(R_1,\cdots, R_L)} (\CL) =\CH_{(R_1,R_1), \cdots, (R_L,R_L)} (\CL) \quad {\rm mod}\, \ 2.
\ee

\sectiono{The conjecture}

\subsection{Review of the conjecture for the colored HOMFLY invariant}

We start by recalling the conjecture of \cite{ov,lmv,lmknot} on the integrality structure of the colored HOMFLY polynomial. We first state the conjecture for knots, 
and then we briefly consider the generalization to links. Notice that these conjectures have now been proved in \cite{lpeng}. 

Let $\CK$ be a knot, and let $\CH_R(\CK)$ be its colored HOMFLY invariant with the coloring $R$. We first define the generating functional 
\be
\label{zh}
Z_{\CH} (v) =\sum_R \CH_R(\CK) s_R(v)
\ee
understood as a formal power series in $s_R(v)$. Here we sum over all possible colorings, including the empty one $R=\cdot$. We also define 
the free energy
\be
F_{\CH}(v) =\log \, Z_{\CH}(v)
\ee
which is also a formal power series. The {\it reformulated HOMFLY invariants} of $\CK$, $f_R(t,\nu)$, are defined through the equation
\be
\label{reformulated}
F_{\CH}(v) = \sum_{d=1}^{\infty} \sum_R {1\over d} f_R(t^d, \nu^d) s_R(v^d). 
\ee
One can easily prove \cite{lmknot} that this equation determines uniquely the reformulated HOMFLY invariants $f_R$ in terms of the colored HOMFLY invariants of 
$\CK$. Explicit formulae for $f_R$ in terms of $\CH_R$ for representations with up to three boxes are listed in \cite{lmknot}. 

If $\ell (R) =\ell (S)$ we define the matrix 
\be
M_{R S}=\sum_{\mu}  {1\over z_{\mu}} \chi_R(C_\mu) \chi_{S}(C_\mu) {\prod_{i=1}^{\ell(\mu)} \bigl( t^{\mu_i} -t^{-\mu_i} \bigr) \over t -t^{-1}} 
\ee
which is zero otherwise. It is easy to show that this matrix is invertible (see for example \cite{lmv,lmknot}). We now define
\be
\label{hattedf}
\hat f_R(t,\nu) =\sum_{S} M_{RS}^{-1} f_S(t, \nu). 
\ee
In principle, $\hat f_R(t,\nu)$ are rational functions, i.e. they belong to the ring $\IQ[t^{\pm 1}, \nu^{\pm 1}]$ with denominators given by products of $t^r-t^{-r}$. However, we have the following 
\begin{conjecture} $\hat f_R(t,\nu) \in z^{-1} \IZ[z^2, \nu^{\pm1}]$, i.e. they have the structure
\be
\hat f_R(t,\nu) =z^{-1} \sum_{g\ge 0} \sum_{Q \in \IZ} N_{R; g,Q} z^{2g} \nu^Q,
\ee
where $N_{R;g,Q}$ are integer numbers and are called the BPS invariants of the knot $\CK$. The sum appearing here is finite, i.e. for a given knot 
and a given coloring $R$, the $N_{R;g,Q}$ vanish except for finitely many values of $g$, $Q$. 
\end{conjecture}

The conjecture can be generalized to links. Let $\CL$ be a link of $L$ components $\CK_1, \cdots, \CK_L$, and let $v_l$, $l=1,\cdots, L$, be formal sets of infinite variables. The subindex $l$ refers here to the $l$-th component of the link, and each $v_l$ has the form $v_l=((v_l)_1, (v_l)_2, \cdots)$. We define 
\be
Z_{\CH}(v_1,\cdots, v_L) =\sum_{R_1, \cdots, R_L} \CH_{R_1, \cdots, R_L}(\CL) s_{R_1}(v_1) \cdots s_{R_L} (v_L)
\ee
as well as the free energy
\be
F_{\CH}(v_1, \cdots, v_L) =\log \, Z_{\CH}(v_1, \cdots, v_L).
\ee
If any of the $R_i$s are given by the trivial coloring $R_i=\cdot$, it is understood that $\CH_{R_1, \cdots, R_L}(\CL)$ is the HOMFLY invariant of the sublink of $\CL$ obtained after removing the corresponding $\CK_i$s. The reformulated invariants are now defined by
\be
F_{\CH}(v_1, \cdots, v_L) = \sum_{d=1}^{\infty} \sum_{R_1, \cdots, R_L} f_{R_1, \cdots, R_L}(t^d, \nu^d) s_{R_1} (v_1^d)\cdots s_{R_L} (v^d_L)
\ee
and
\be
\hat f_{R_1, \cdots, R_L} (t,\nu) =\sum_{S_1, \cdots, S_L} M_{R_1 S_1 }^{-1}\cdots M_{R_L S_L }^{-1} f_{S_1, \cdots, S_L}(t, \nu). 
\ee
\begin{remark} Notice that, for the fundamental representation, 
\be
\hat f_{\tableau{1}, \cdots, \tableau{1}} (t,\nu) =f_{\tableau{1}, \cdots, \tableau{1}} (t,\nu).
\ee
\end{remark}

We can now state the conjecture for links.
\begin{conjecture} $\hat f_{R_1, \cdots, R_L} (t,\nu) \in z^{L-2} \IZ[z^2, \nu^{\pm1}]$, i.e. they have the structure
\be
\hat f_{R_1, \cdots, R_L} (t,\nu) =z^{L-2} \sum_{g\ge 0} \sum_{Q \in \IZ} N_{R_1, \cdots, R_L;g,Q} z^{2g} \nu^Q.
\ee
\end{conjecture}

These conjectures also hold for framed knots and links \cite{mv}. 

\subsection{The conjecture for the colored Kauffman invariant}

We will first state the conjecture for knots. Let $\CK$ be an oriented knot, and let $\CH_{(R,S)}$ be its HOMFLY invariant in the composite 
representation $(R,S)$. The {\it composite} invariant of the knot $\CK$, colored by the representation $R$, and denoted by $\CR_R$, is given by
\be
\CR_R (\CK)=\sum_{R_1, R_2} N^R_{R_1 R_2} \CH_{(R_1, R_2)}(\CK)
\ee
where $N_{R_1 R_2}^R$ are the Littlewood--Richardson coefficients defined by (\ref{lr}). Notice that, due to (\ref{hcomin}), this invariant is independent 
of the choice of orientation of the knot, and it is therefore and invariant of unoriented knots. 

\begin{example} 
We give some simple examples of the composite invariant for colorings with up to two boxes:
\be
\ba
\CR_{\tableau{1}} &=2 \CH_{\tableau{1}},\\
\CR_{\tableau{2}}&=2 \CH_{\tableau{2}} + \CH_{(\tableau{1},\tableau{1})},\\
\CR_{\tableau{1 1}}&=2 \CH_{\tableau{1 1}} + \CH_{(\tableau{1},\tableau{1})}.
\ea
\ee
\end{example}
Using these invariants we define the generating functionals
\be
\label{zr}
Z_{\CR}(v)=\sum_R \CR_R(\CK) s_R(v), \qquad F_{\CR}(v) =\log \, Z_{\CR}(v).
\ee
We also define the generating functionals for colored Kauffman invariants of $\CK$ as
\be
\label{zg}
Z_{\CG}(v)=\sum_R \CG_R(\CK) s_R(v),\qquad F_{\CG}(v) =\log \, Z_{\CG}(v).
\ee
This allows us to define two sets of reformulated invariants, $h_R$ and $g_R$, as follows. The $h_R$ are defined by a relation identical to 
(\ref{reformulated}), 
\be
F_{\CR}(v) =\sum_{d=1}^{\infty} \sum_R {1\over d} h_R(t^d, \nu^d) s_R(v^d)
\ee
while the $g_R$ are defined by 
\be
\label{grdef}
F_{\CG}(v) -{1\over 2} F_{\CR}(v)=\sum_{d \, {\rm odd}} \sum_R {1\over d} g_R(t^d, \nu^d) s_R(v^d).
\ee
Here the sum over $d$ is over all 
positive {\it odd} integers. $h_R$ can be explicitly obtained in terms of colored HOMFLY 
invariants for composite representations, while the $g_R$ can be written in terms of 
these invariants and the colored Kauffman invariants. 

\begin{example} We list here explicit expressions for the reformulated invariants $g_R$ of a knot, where $R$ is a representation of up to three boxes. We have
\be
\label{gknots}
\ba
g_{\tableau{1}}&=\CG_{\tableau{1}}-{\cal H}_{\tableau{1}}, \\
g_{\tableau{2}}&=\CG_{\tableau{2}}-{1\over 2}\CG_{\tableau{1}}^2 -{\cal H}_{\tableau{2}} + {\cal H}_{\tableau{1}}^2
-{1\over 2}\CH_{(\tableau{1}, \tableau{1})}, \\
g_{\tableau{1 1}}&=\CG_{\tableau{1 1}}- {1\over 2}\CG_{\tableau{1}}^2 -{\cal H}_{\tableau{1 1}} +
{\cal H}_{\tableau{1}}^2-{1\over 2}\CH_{(\tableau{1},\tableau{1})}, 
\ea
\ee
as well as
\be
\ba
\label{gknotstwo}
g_{\tableau{3}}&=\CG_{\tableau{3}}-\CG_{\tableau{2}} \CG_{\tableau{1}} +{1 \over 3} \CG^3_{\tableau{1}}-{1\over 3} g_{\tableau{1}}(t^3, \nu^3)\\
&-{\cal H}_{\tableau{3}}
-{\cal H}_{(\tableau{2},\tableau{1})}+2 \CH_{\tableau{2}} \CH_{\tableau{1}} +\CH_{(\tableau{1}, \tableau{1})} \CH_{\tableau{1}} - {4\over 3} \CH_{\tableau{1}}^3,\\
g_{\tableau{2 1}}&=\CG_{\tableau{2 1}}-\CG_{\tableau{2}} \CG_{\tableau{1}} -\CG_{\tableau{1 1}} \CG_{\tableau{1}}+{2 \over 3} \CG^3_{\tableau{1}}+{1\over 3} g_{\tableau{1}}(t^3, \nu^3)\\
&-{\cal H}_{\tableau{2 1}} -{\cal H}_{(\tableau{2},\tableau{1})} -{\cal H}_{(\tableau{1 1},\tableau{1})} 
+2 {\cal H}_{\tableau{2}} {\cal H}_{\tableau{1}}+2 {\cal H}_{\tableau{1 1}} {\cal H}_{\tableau{1}}+2\CH_{(\tableau{1}, \tableau{1})} \CH_{\tableau{1}} - {8\over 3} \CH_{\tableau{1}}^3 ,\\
g_{\tableau{1 1 1}}
&=\CG_{\tableau{1 1 1}}-\CG_{\tableau{1 1}} \CG_{\tableau{1}} +{1 \over 3} \CG^3_{\tableau{1}}-{1\over 3} g_{\tableau{1}}(t^3, \nu^3)\\
&-{\cal H}_{\tableau{1 1 1}}
-{\cal H}_{(\tableau{1 1},\tableau{1})}+2 \CH_{\tableau{1 1}} \CH_{\tableau{1}} +\CH_{(\tableau{1}, \tableau{1})} \CH_{\tableau{1}} - {4\over 3} \CH_{\tableau{1}}^3.
\ea
\ee
\end{example}

The invariants $\hat h_R$, $\hat g_R$ are defined by a relation identical to (\ref{hattedf}), 
\be
\hat h_R(t,\nu) =\sum_{S} M_{RS}^{-1} h_S(t, \nu), \qquad \hat g_R(t,\nu) =\sum_{S} M_{RS}^{-1} g_S(t, \nu). 
\ee
Like before, $\hat h_R(t,\nu)$ and $\hat g_R (t,\nu)$ belong in principle to the ring $\IQ[t^{\pm 1}, \nu^{\pm 1}]$ with denominators given by products of $t^r-t^{-r}$. The 
conjecture for the colored Kauffman polynomial states an integrality property similar to the one we stated for the colored HOMFLY invariant.
\begin{conjecture} \label{myconj} We have that 
\be
\hat h_R(t,\nu) \in z^{-1} \IZ[z^2, \nu^{\pm1}], \qquad \hat g_R(t,\nu) \in  \IZ[z, \nu^{\pm1}] 
\ee
i.e. they have the structure
\be
\ba
\hat h_R(t,\nu) &=z^{-1} \sum_{g\ge 0} \sum_{Q \in \IZ} N^{c=0}_{R; g,Q} z^{2g} \nu^Q, \\
\hat g_R (t,\nu)&=\sum_{g\ge 0} \sum_{Q \in \IZ} \Bigl( N^{c=1}_{R; g,Q} z^{2g} \nu^Q + N^{c=2}_{R; g,Q} z^{2g+1} \nu^Q \Bigr), 
\ea
\ee
where $N_{R;g,Q}^{c=0,1,2}$ are integers. 
\end{conjecture}

Again, there is a generalization to links as follows. Let $\CL$ be an unoriented link, and pick an arbitrary orientation. We define the 
composite invariant of $\CL$ as
\be
\label{linkoR}
\CR_{R_1, \cdots, R_L} (\CL)=\sum_{U_1, V_1, \cdots, U_L, V_L} N^{R_1}_{U_1 V_1} \cdots N^{R_L}_{U_L V_L} \CH_{(U_1, V_1), \cdots, (U_L, V_L)}(\CL).
\ee
Due to (\ref{reversal}), this invariant does not depend on the choice of orientation of $\CL$, and it is therefore an invariant of unoriented links. We further define 
the generating functionals
\be
\ba
Z_{\CR}(v_1,\cdots, v_L) &=\sum_{R_1, \cdots, R_L} \CR_{R_1, \cdots, R_L}(\CL) s_{R_1}(v_1) \cdots s_{R_L} (v_L), \\
Z_{\CG}(v_1,\cdots, v_L) &=\sum_{R_1, \cdots, R_L} \CG_{R_1, \cdots, R_L}(\CL) s_{R_1}(v_1) \cdots s_{R_L} (v_L),
\ea
\ee
as well as the free energies
\be
F_{\CR}(v_1, \cdots, v_L) =\log \, Z_{\CR}(v_1, \cdots, v_L), \qquad F_{\CG}(v_1, \cdots, v_L) =\log \, Z_{\CG}(v_1, \cdots, v_L).
\ee
The reformulated invariants $h_{R_1, \cdots, R_L}$, $g_{R_1, \cdots, R_L}$ are now defined by
\be
F_{\CR}(v_1, \cdots, v_L) = \sum_{d=1}^{\infty} \sum_{R_1, \cdots, R_L} h_{R_1, \cdots, R_L}(t^d, \nu^d) s_{R_1} (v_1^d)\cdots s_{R_L} (v^d_L)
\ee
and
\be
F_{\CG}(v_1, \cdots, v_L) ) -{1\over 2} F_{\CR}(v_1, \cdots, v_L)=\sum_{d \, {\rm odd}} \sum_{R_1, \cdots, R_L} g_{R_1, \cdots, R_L}(t^d, \nu^d) s_{R_1} (v_1^d)\cdots s_{R_L} (v^d_L).
\ee
Finally, the ``hatted" invariants are defined by the relation
\be
\ba
\hat h_{R_1, \cdots, R_L} (t,\nu) &=\sum_{S_1, \cdots, S_L} M_{R_1 S_1 }^{-1}\cdots M_{R_L S_L }^{-1} h_{S_1, \cdots, S_L}(t, \nu),\\
\hat g_{R_1, \cdots, R_L} (t,\nu) &=\sum_{S_1, \cdots, S_L} M_{R_1 S_1 }^{-1}\cdots M_{R_L S_L }^{-1} g_{S_1, \cdots, S_L}(t, \nu). 
\ea\ee
\begin{example}
For links $\CL$ of two components $\CK_1$, $\CK_2$ we have
\be
\label{glinks}
\ba
g_{\tableau{1} \tableau{1}} (\CL)&=\CG_{\tableau{1}, \tableau{1}}(\CL) -\CG_{\tableau{1}}(\CK_1) \CG_{\tableau{1}}(\CK_2)-\CH_{\tableau{1}, \tableau{1}}(\CL)-\CH_{\tableau{1}, \tableau{1}}({\overline \CL}) + 
2 \CH_{\tableau{1}}(\CK_1) \CH_{\tableau{1}}(\CK_2),\\
g_{\tableau{2} \tableau{1}} (\CL)&=\CG_{\tableau{2}, \tableau{1}}(\CL)- \CG_{\tableau{1}, \tableau{1}}(\CL)\CG_{\tableau{1}}(\CK_1)-\CG_{\tableau{2}}(\CK_1) \CG_{\tableau{1}}(\CK_2)+ \CG_{\tableau{1}}(\CK_1)^2 \CG_{\tableau{1}}(\CK_2)\\
&-\CH_{\tableau{2}, \tableau{1}}(\CL)-\CH_{\tableau{2}, \tableau{1}}({\overline \CL})-\CH_{(\tableau{1},\tableau{1}),\tableau{1}}(\CL)+ 2\bigl(\CH_{\tableau{1}, \tableau{1}}(\CL)+\CH_{\tableau{1}, \tableau{1}}({\overline \CL})\bigr)\CH_{\tableau{1}}(\CK_1)\\
&+2 \CH_{\tableau{2}} (\CK_1)\CH_{\tableau{1}}(\CK_2)+ \CH_{(\tableau{1},\tableau{1})}(\CK_1) \CH_{\tableau{1}}(\CK_2) -4 \CH_{\tableau{1}}(\CK_1)^2 \CH_{\tableau{1}}(\CK_2),\\
g_{\tableau{1 1} \tableau{1}} (\CL)&=\CG_{\tableau{1 1}, \tableau{1}}(\CL)- \CG_{\tableau{1}, \tableau{1}}(\CL)\CG_{\tableau{1}}(\CK_1)-\CG_{\tableau{2}}(\CK_1) \CG_{\tableau{1}}(\CK_2)+ \CG_{\tableau{1}}(\CK_1)^2 \CG_{\tableau{1}}(\CK_2)\\
&-\CH_{\tableau{1 1}, \tableau{1}}(\CL)-\CH_{\tableau{1 1}, \tableau{1}}({\overline \CL})-\CH_{(\tableau{1},\tableau{1}),\tableau{1}}(\CL)+ 2\bigl(\CH_{\tableau{1}, \tableau{1}}(\CL)+\CH_{\tableau{1}, \tableau{1}}({\overline \CL})\bigr)\CH_{\tableau{1}}(\CK_1)\\
&+2 \CH_{\tableau{1 1}} (\CK_1)\CH_{\tableau{1}}(\CK_2)+ \CH_{(\tableau{1},\tableau{1})}(\CK_1) \CH_{\tableau{1}}(\CK_2) -4 \CH_{\tableau{1}}(\CK_1)^2 \CH_{\tableau{1}}(\CK_2).
\ea
\ee
In these equations, ${\overline {\CL}}$ is the link obtained from $\CL$ by inverting the orientation of one of its components. 
\end{example}

\begin{figure}[!ht]
\leavevmode
\begin{center}
\epsfysize=1.3cm
\epsfbox{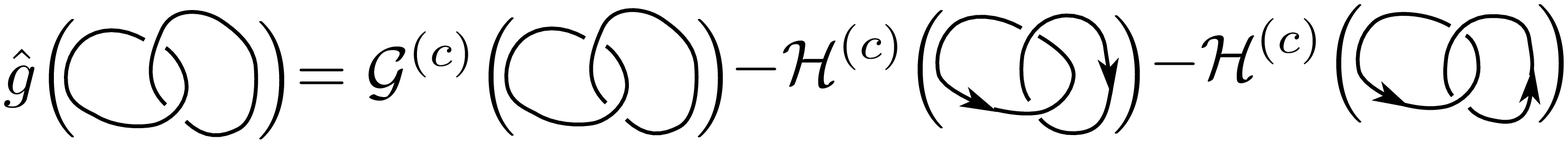}
\end{center}
\caption{The reformulated invariant of an unoriented link $\CL$, $\hat g_{R_1, \cdots R_L}(\CL)$, involves the Kauffman invariant of $\CL$, together with the HOMFLY invariants of all possible choices of orientations for the components of the link. Here we illustrate it for the fundamental representation and for a two-component link.}
\label{combinatorics}
\end{figure}

\begin{example} For general links of $L$ components it is easy to write down a general formula for $g_{\tableau{1} , \cdots, \tableau{1}}(\CL)$. 
We first define the connected Kauffman invariant of a link $\CL$ as the term 
   multiplying $s_{\tableau{1}}(v_1)\cdots s_{\tableau{1}}(v_L)$ in the expansion of $F_\CG (v_1, \cdots, v_L)$. It is given by 
\be
\label{connlink}
\CG^{(c)}(\CL)=\CG(\CL)-\sum_{j=1}^L \CG (\CK_j)\CG (\CL_j) +\cdots
\ee
where the link $\CL_j$ is obtained from $\CL$ by removing the $j$-th component. Further corrections involve all possible sublinks of $\CL$, and 
the combinatorics appearing in the formula is the same one that appears in the calculation of the cumulants of a probability distribution. 
A similar definition gives the connected HOMFLY invariant of a link, $\CH^{(c)}(\CL)$, which was studied in detail in \cite{lmv,lm}. 
We now consider all possible oriented links that can be obtained from an unoriented link $\CL$ 
of $L$ components by choosing different orientations in their component knots. In principle there are $2^L$ oriented links that can be obtained in this way, but they can be grouped in pairs that differ in an overall reversal of orientation, and therefore lead to the same 
HOMFLY invariant. We conclude that there are $2^{L-1}$ different links which differ in the relative orientation of their components and have {\it a priori} different HOMFLY invariants. We will denote these links 
by ${\overline \CL}_\alpha$, where $\alpha=1, \cdots, 2^{L-1}$. Using (\ref{reversal}), it is easy to see that the oriented invariant (\ref{linkoR}) for $R_1=\cdots=R_L=\tableau{1}$ 
involves the sum over all possible orientations of the link, and we have 
\be
\CR_{\tableau{1}, \cdots, \tableau{1}} (\CL)= 2 \sum_{\alpha=1}^{2^{L-1}} \CH ({\overline \CL}_{\alpha}). 
\ee
The reformulated invariant
$ g_{\tableau{1} , \cdots, \tableau{1}}(\CL)$ is then given by 
 \be
 \label{hatglink}
g_{\tableau{1} , \cdots, \tableau{1}}(\CL) =\CG^{(c)}(\CL) -\sum_{\alpha=1}^{2^{L-1}} \CH^{(c)}({\overline \CL}_\alpha).
\ee
In general, the reformulated invariant of an unoriented link $\CL$,  $\hat g_{R_1, \cdots R_L}(\CL)$, 
involves colored Kauffman invariants of $\CL$, together with colored HOMFLY invariants of all possible choices of orientations 
of the link. This is an important feature of the reformulated invariants, and we illustrate it graphically for a two-component link in \figref{combinatorics}. The fact that one has to consider all possible orientations of the unoriented link bears some resemblance to 
Jaeger's model for the Kauffman polynomial in terms of the HOMFLY polynomial (see for example \cite{kauffmanbook}, pp. 219-222), and it has appeared before in the context of the 
Kauffmann invariant in \cite{lmtformula}. 
\end{example}

We can now state our conjecture for the Kauffman invariant of links.
\begin{conjecture} 
We have that
\be
 \hat h_{R_1, \cdots, R_L} (t,\nu) \in z^{L-2} \IZ[z^2, \nu^{\pm1}], \qquad \hat g_{R_1, \cdots, R_L} (t,\nu) \in z^{L-1} \IZ[z, \nu^{\pm1}]
\ee
 i.e. they have the structure
\be
\ba
\hat h_{R_1, \cdots, R_L} (t,\nu) &=z^{L-2} \sum_{g\ge 0} \sum_{Q \in \IZ} N^{c=0}_{R_1, \cdots, R_L;g,Q} z^{2g} \nu^Q, \\
\hat g_{R_1, \cdots, R_L}  (t,\nu)&=z^{L-1}\sum_{g\ge 0} \sum_{Q \in \IZ} \Bigl( N^{c=1}_{R_1, \cdots, R_L; g,Q} z^{2g} \nu^Q + N^{c=2}_{R_1, \cdots, R_L; g,Q} z^{2g+1} \nu^Q \Bigr).
\ea
\ee
\end{conjecture}

\begin{remark} It follows from this conjecture that 
\be
h_{R_1, \cdots, R_L}  \in z^{L-2} \IZ[t^{\pm 1}, \nu^{\pm1}], \qquad g_{R_1, \cdots, R_L}  \in  z^{L-1} \IZ[t^{\pm 1}, \nu^{\pm1}].
\ee
As in the colored HOMFLY case, the conjecture is supposed to hold as well for framed knots and links. 
\end{remark}
\sectiono{Evidence}

\subsection{Direct computations}

In this section we provide some evidence for our conjectures concerning the colored Kauffman invariant of knots and links. 
The first type of evidence follows from direct computation of the invariants $\hat h_{R_1, \cdots, R_L}$, 
$\hat g_{R_1, \cdots, R_L}$ for simple knots and links and for representations with small number of boxes. 

\begin{example} The simplest example is of course the unknot. The colored HOMFLY and Kauffman invariants are just quantum dimensions. For the standard framing one finds 
that the only nonvanishing $\hat h_R$, $\hat g_R$ are 
\be
\ba
\hat h_{\tableau{1}} &= 2(\nu -\nu^{-1})z^{-1}, \\
\hat h_{\tableau{2}}&=-z^{-1}, \\ 
\hat h_{\tableau{1 1}}&=-z^{-1}, \\
\hat g_{\tableau{1}}&=1.
\ea
\ee
Although we have only computed the reformulated invariants up to four boxes, we conjecture that the $\hat h_R$, $\hat g_R$ vanish for all remaining representations. Of course, this has the structure 
predicted by our conjecture. 
\end{example}

We now consider more complicated examples. As in \cite{lm,lmv}, a useful testing ground are torus knots and links, since 
for them one can write down general expressions for the colored invariants in any representation. Torus knots are labelled by two coprime integers $n,m$, and we will 
denote them by $\CK_{n,m}$. Torus links are labelled by two integers, and their g.c.d. is the number of components of the link, $L$. We will denote a 
torus link by $\CL_{L n, Lm}$, where $n,m$ are coprime. Explicit formulae for the HOMFLY invariant of a torus knot $\CK_{n,m}$, colored by a representation $R$, can be obtained in many ways. In the context of Chern--Simons theory, one can use for example the formalism of knot operators of \cite{lllr} to write down general expressions \cite{lm}. In fact \cite{stevan}, 
one can obtain formulae in the knot operator formalism which are 
much simpler than those presented in \cite{lm} and make contact with the elegant result derived in \cite{lz} by using Hecke algebras. The formula one 
obtains is simply 
\be
\label{homknots}
\CH_R (\CK_{n,m})=t^{nm C_R}
\sum_U c_{n;R}^U \, t^{-{m\over n}  C_U} \, 
 {\rm dim}_q \, U
\ee
where $c_{n;R}^U$ is defined in (\ref{cnr}). Of course one has to set $t^N=\nu$. This formula is also valid for composite representations, which after all are just a special type of representations of 
${\rm U}(N)$. The generalization to torus links is immediate, as noticed in \cite{lmv}, and the invariant for $\CL_{Ln,Lm}$ is given by 
\be\label{homlinks}
\CH_{R_1, \cdots, R_L} (\CL_{Ln,Lm})=\sum_S N^S_{R_1, \cdots, R_L} t^{mn\bigl( C_S -\sum_{j=1}^L C_{R_j}\bigr)} \CH_S (\CK_{n,m}).
\ee
This expression is also valid for composite representations, but one has to use the appropriate Littlewood--Richardson coefficients (as computed in for example \cite{jap}). 

\begin{example} By using these formulae one obtains, for the trefoil knot, 
\be
\ba
\CH_{(\tableau{1},\tableau{1})}(\CK_{2,3})={\rm dim}_q \, (\tableau{1},\tableau{1}) \, \Bigl\{ & 4\nu^4 - 4 \nu^6 +\nu^8 + z^2\bigl(  4\nu^4 - 7 \nu^6 +2 \nu^8 +\nu^{10}\bigr) \\ 
&+ 
z^4  \bigl(\nu^4 - 2 \nu^6 +\nu^8 \bigr) \Bigr\},
\ea
\ee 
while for the Hopf link we have for example
\be
\CH_{(\tableau{1},\tableau{1}), (\tableau{1},\cdot)}(\CL_{2,2})=\bigl( {\rm dim}_q \, (\tableau{1},\tableau{1}) \bigr) \bigl({\rm dim}_q \, \tableau{1} \bigr) (1+z^2).
\ee
\end{example}

\begin{remark} In the case of the Hopf link, a general 
expression for $\CH_{(R_1, S_1), (R_2, S_2)}$ in terms of the topological vertex \cite{akmv} can be read from the results for the ``covering contribution" in \cite{bfmtwo}. This expression has reappeared in other studies of topological string theory, see \cite{anv,kanno}. Particular cases have been computed by using skein theory in \cite{hmtwo}.  
\end{remark}

One also needs to compute the colored Kauffman invariants of torus knots and links. Very likely, the expression (\ref{homknots}) generalizes to the the Kauffman case by using the group theory data for ${\rm SO}(N)/{\rm Sp}(N)$, but we have used the expression presented in \cite{bfmtwo} for torus knots of the type $(2, 2m+1)$, based on the approach of \cite{rkg}. With these ingredients it is straightforward to compute the 
reformulated invariants $\hat g_R$ for torus knots, although the expressions quickly become quite complicated. 
We have verified the conjecture for various framed torus knots and links and representations with up to four boxes. 

\begin{example} For the trefoil knot in the standard framing one finds
\be
\ba
\hat g_{\tableau{2}}&=-21 \nu ^{11}+79 \nu ^9-111 \nu ^7+69 \nu ^5-16 \nu ^3 +21z  \left(\nu ^{12}-3 \nu ^{10}+3 \nu ^8-\nu^6\right) \\
&+z^2\left( -70 \nu ^{11}+251 \nu ^9-307 \nu ^7+146 \nu ^5-20 \nu ^3\right) +7 z^3 \left(10 \nu ^{12}-33 \nu ^{10}+33 \nu ^8-10
   \nu ^6\right) \\ &-2z^4 \left(42 \nu ^{11}-165 \nu ^9+183 \nu ^7-64 \nu ^5+4 \nu ^3\right)+14 z^5 \left(6 \nu ^{12}-23 \nu
   ^{10}+23 \nu ^8-6 \nu ^6\right)\\
   &+z^6\left(-45 \nu ^{11}+220 \nu ^9-230 \nu ^7+56 \nu ^5-\nu ^3\right) +3z^7 \left(15 \nu ^{12}-73 \nu
   ^{10}+73 \nu ^8-15 \nu ^6\right)\\&+z^8 \left(-11 \nu ^{11}+78 \nu ^9-79 \nu ^7+12 \nu ^5 \right)+z^9\left(  -11 \nu ^{11}+78 \nu ^9-79 \nu ^7+12 \nu ^5\right)\\
& +z^{10}\left(  -\nu ^{11}+14 \nu ^9-14 \nu ^7+\nu ^5 \right) + z^{11}\left(\nu ^{12}-14 \nu ^{10}+14 \nu ^8-\nu ^6 \right)\\
&+ z^{12} \left(\nu ^9-\nu
   ^7\right) +z^{13} \left(\nu ^8-\nu ^{10}\right),\\
   \hat g_{\tableau{1 1}}&=-15 \nu ^{11}+53 \nu ^9-69 \nu ^7+39 \nu ^5-8 \nu ^3 +15 z \left(\nu ^{12}-3 \nu ^{10}+3 \nu ^8-\nu
   ^6\right)\\
   &+z^2\left(-35 \nu ^{11}+126 \nu ^9-146 \nu ^7+61 \nu ^5-6 \nu ^3\right)+ 5 z^3 \left(7 \nu ^{12}-24 \nu ^{10}+24 \nu ^8-7 \nu
   ^6\right)  \\
   &+z^4 \left(-28 \nu ^{11}+120 \nu ^9-128 \nu ^7+37 \nu ^5-\nu ^3\right) + 7 z^5 \left(4 \nu ^{12}-17 \nu ^{10}+17 \nu ^8-4 \nu ^6\right)\\
   &+z^6\left( -9 \nu ^{11}+55 \nu ^9-56 \nu ^7+10 \nu ^5\right) +z^7\left( 9 \nu ^{12}-55 \nu ^{10}+55 \nu ^8-9 \nu ^6\right)\\
 & +z^8\left(  -\nu
   ^{11}+12 \nu ^9-12 \nu ^7+\nu ^5\right) + z^9\left( \nu ^{12}-12 \nu ^{10}+12 \nu ^8-\nu ^6\right) \\
   &+z^{10} \left( \nu ^9-\nu ^7\right)+ z^{11}\left( \nu ^8-\nu
   ^{10}\right).
   \ea
   \ee
 From these expressions one can read the BPS invariants $N_{\tableau{2};g,Q}^{c=1,2}$ and $N_{\tableau{1 1};g,Q}^{c=1,2}$. In \cite{bfmtwo} the invariants with $c=1$ were obtained by exploited parity properties of the Kauffman invariant, 
 but the $c=2$ invariants were not determined. Notice that our convention for the matrix $M_{RS}$ is different from the one in \cite{bfmtwo}, so in order to compare with the results for $c=1$ presented in \cite{bfmtwo} one has to change $N_{R;g,Q} \rightarrow (-1)^{\ell(R)-1} N_{R^T;g, Q}$.
 
 \end{example}
\begin{example}  For the Hopf link one finds
 \be
 \ba
 \hat g_{\tableau{1}, \tableau{1}}&=z (\nu-\nu^{-1}),\\
 \hat g_{\tableau{2},\tableau{1}}&=z(\nu^2-1),\\
  \hat g_{\tableau{1 1},\tableau{1}}&=z(\nu^{-2}-1).
  \ea
  \ee
\end{example}
\subsection{General predictions for knots}

We now discuss general predictions of our conjecture. 
We will see that it makes contact with well-known properties of the Kauffman invariant, and that it makes some simple, new predictions for the 
structure of the Kauffman invariant of links. We start by discussing general predictions for knots.

From their definition (\ref{hinv}), (\ref{kinv}) we can write
\be
\ba
\CG(\CK)&=\Bigl( 1+ {\nu -\nu^{-1} \over z} \Bigr) \sum_{i\ge 0} k_i^{\CK}(\nu) z^{i}, \\
\CH(\CK)&={\nu -\nu^{-1} \over z} \sum_{i\ge 0} p_i^{\CK}(\nu) z^{2i}.
\ea
\ee
According to our conjecture, $\hat g_{\tableau{1}}=g_{\tableau{1}}$ has no terms in $z^{-1}$, therefore one must have
\be
\label{pkknots}
k_0^{\CK}(\nu)=p_0^{\CK}(\nu),
\ee
which is (\ref{pk}) in the case of knots. Therefore, the equality of the lowest order terms of the HOMFLY and Kauffman polynomials is a simple consequence of our conjecture. 
This was already noticed in \cite{bfmtwo}.

We now consider the reformulated polynomial $g_R$ for representations with two boxes. Our conjecture implies that this quantity belongs to $z^{-1} \IZ[\nu^{\pm1}, t^{\pm1}]$. By looking at the 
definition of $g_{\tableau{2}}(\CK)$, $g_{\tableau{1 1}}(\CK)$ in terms of colored Kauffman and HOMFLY invariants, we see that the only possible term which might spoil integrality is
\be
{1\over 2} (\CG(\CK)^2 + \CH_{(\tableau{1},\tableau{1})}(\CK)).
\ee
Therefore our conjecture implies that 
\be
\CG(\CK)^2\equiv  \CH_{(\tableau{1},\tableau{1})}(\CK) \quad {\rm mod}\, 2.
\ee
This is precisely Rudolph's theorem  (\ref{rudolphth}) for knots. 

Very likely, our integrality conjecture also leads to the generalization 
of Rudolph's theorem due to Morton and Ryder (\ref{mrgen}), although the combinatorics becomes more involved. 
As an example, we will briefly show how to derive (\ref{mrgen}) in the case of knots ($L=1$) and with $R=\tableau{2}$. To do this, we look at $g_{\tableau{4}}$, which is given by
\be
\ba
g_{\tableau{4}}&= \CG_{\tableau{4}} -\CG_{\tableau{1}}  \CG_{\tableau{3}}  -{1\over 2}\CG^2_{\tableau{2}}  + {1\over 2} \CG_{\tableau{1}}^2 \CG_{\tableau{2}} -{1\over 4} \CG^4_{\tableau{1}} \\
&-{1\over 2} \biggl( \CR_{\tableau{4}} -\CR_{\tableau{1}}  \CR_{\tableau{3}}  -{1\over 2}\CR^2_{\tableau{2}}  + {1\over 2} \CR_{\tableau{1}}^2 \CR_{\tableau{2}} -{1\over 4} \CR^4_{\tableau{1}} \biggr).
\ea
\ee
Most terms in the r.h.s. are manifestly elements in $\IZ[t^{\pm 1}, \nu^{\pm1}]$ with denominators given by products of $t^r-t^{-r}$. The only possible source for rational coefficients is the term 
\be
-{1\over 2} \Bigl( \CG_{\tableau{2}}^2 + \CH_{(\tableau{2}, \tableau{2})}\Bigr) -{1\over 4} \Bigl( \CG_{\tableau{1}}^4 -\CH_{(\tableau{1}, \tableau{1})}^2\Bigr). 
\ee
However, the last two terms inside the bracket are also in $\IZ[t^{\pm 1}, \nu^{\pm1}]$ thanks to Rudolph's theorem, and we conclude that integrality of $g_{\tableau{4}}$ requires
\be
\CG_{\tableau{2}}^2 \equiv  \CH_{(\tableau{2}, \tableau{2})} \qquad {\rm mod}\, \, 2. 
\ee
This is Morton--Ryder's theorem (\ref{mrgen}) for a knot colored by $R=\tableau{2}$. It seems likely that the general case of their theorem, for an arbitrary representation $R$, 
follows from integrality of $g_S$, where $S \in R\otimes R$.

\subsection{General predictions for links}
Let us now consider two-component links. When both components are colored by $\tableau{1}$, the HOMFLY and Kauffman invariants have the form
\be
\ba
\CG(\CL)&=\Bigl( 1+ {\nu -\nu^{-1} \over z} \Bigr) \sum_{i\ge 0} k_i^{\CL}(\nu) z^{i-1}, \\
\CH(\CL)&={\nu -\nu^{-1} \over z} \sum_{i\ge 0} p_i^{\CL}(\nu) z^{2i-1}.
\ea
\ee
Our conjecture says that $g_{\tableau{1}, \tableau{1}}(\CL)$ belongs to $\IZ[\nu^{\pm1}, z]$, i.e. it has no negative powers of $z$. From its explicit definition in terms of the HOMFLY and Kauffman polynomials of $\CL$ we find 
that this condition leads to three different relations. The first one is
\be
\label{kzeros}
k_0^{\CL}(\nu)=(\nu-\nu^{-1}) k^{\CK_1}_0(\nu) k^{\CK_2}_0(\nu).
\ee
The conjecture in the case of HOMFLY leads to a similar relation \cite{lmv}
\be
\label{pzeros}
p_0^{\CL}(\nu)=(\nu-\nu^{-1}) p^{\CK_1}_0(\nu) p^{\CK_2}_0(\nu).
\ee
Notice that, due to (\ref{pkknots}), we also have from (\ref{pzeros}) and (\ref{kzeros}), that
\be
p_0^{\CL}(\nu)=k_0^{\CL}(\nu)
\ee
for links of two components. 
The second relation determines the second coefficient of the Kauffman polynomial of a link as 
\be
\label{kone}
 k_1^{\CL}(\nu)=k^{\CK_1}_0(\nu) k^{\CK_2}_0(\nu) +(\nu-\nu^{-1}) \bigl( k^{\CK_1}_0(\nu) k^{\CK_2}_1(\nu) +k^{\CK_1}_1(\nu) k^{\CK_2}_0(\nu) \bigr).
 \ee
 Finally, the third relation gives an equation for $k_2^{\CL}(\nu)$, 
 \be
 \ba
 \label{ktwo}
  k_2^{\CL}(\nu)&=p_1^{\CL}(\nu) + p_1^{{\overline \CL}}(\nu) -2 \bigl(p^{\CK_1}_0(\nu) p^{\CK_2}_1(\nu) +p^{\CK_1}_1(\nu) p^{\CK_2}_0(\nu)\bigr) \\
  &+k^{\CK_1}_0(\nu) k^{\CK_2}_1(\nu) +k^{\CK_1}_1(\nu) k^{\CK_2}_0(\nu)\\
  & +(\nu-\nu^{-1})\bigl( k^{\CK_1}_0 (\nu) k^{\CK_2}_2(\nu) +k^{\CK_1}_2 (\nu) k^{\CK_2}_0(\nu)+k^{\CK_1}_1(\nu) k^{\CK_2}_1(\nu)\bigr).
  \ea
  \ee

   These results can be easily generalized to a general link $\CL$ with $L$ components $\CK_j$, $j=1, \cdots, L$, as follows. If we calculate the connected invariants of the link from their definitions in terms of invariants of sublinks, we obtain an expression of the form
\be
\ba
\CG^{(c)}(\CL)&=\Bigl( 1+ {\nu -\nu^{-1} \over z} \Bigr) \sum_{i\ge 0} k_i^{(c),\CL}(\nu) z^{i+1-L}, \\
\CH^{(c)}(\CL)&={\nu -\nu^{-1} \over z} \sum_{i\ge 0} p_i^{(c),\CL}(\nu) z^{2i+1-L},
\ea
\ee
where $p_i^{(c),\CL}(\nu)$, $k_i^{(c),\CL}(\nu)$ can be obtained in terms of the polynomials $p_i^{(c),\CL'}(\nu)$, $p_i^{(c),\CL'}(\nu)$ of the different sublinks of $\CL$, $\CL'\subset \CL$. For example, 
\be
p_0^{(c),\CL}(\nu)=p_0^{\CL}(\nu)-(\nu -\nu^{-1})^{L-1}\prod_{j=1}^{L-1} p_0^{\CK_j}(\nu). 
\ee
The conjecture of \cite{ov,lmv,lmknot} for the colored HOMFLY invariant implies in particular that the connected HOMFLY invariant belongs to $z^{L-2} \IZ[z^2,\nu^{\pm1}]$. This leads to \cite{lmv,lmknot}
\be
\label{homflylink}
p_0^{(c),\CL}(\nu)=\cdots =p_{L-2}^{(c),\CL}(\nu)=0
\ee
for any link $\CL$. The fact that $p_0^{(c),\CL}(\nu)=0$ is a result of Lickorish and Millett \cite{limi}, and the vanishing of $p_{1,2}^{(c),\CL}(\nu)$ has been proved in \cite{km}. 

Our conjecture for the colored Kauffman implies that (\ref{hatglink}) belongs to $z^{L-1} \IZ[z,\nu^{\pm1}]$. This gives the relations
\be
\label{genlinkvanish}
k_0^{(c),\CL}(\nu)=\cdots =k_{2L-3}^{(c),\CL}(\nu)=0
\ee
as well as 
\be
\label{genlinkorientation}
k_{2L-2}^{(c),\CL}(\nu)=\sum_{\alpha=1}^{2^{L-1}} p_{L-1}^{(c),\overline {\CL}_\alpha}(\nu).
\ee
The relations (\ref{genlinkvanish}) generalize (\ref{kzeros}) and (\ref{kone}), while (\ref{genlinkorientation}) generalizes (\ref{ktwo}) to any link. The equality (\ref{pk}) for any link now follows from the vanishing of $p_0^{(c),\CL}(\nu)$, $k_0^{(c),\CL}(\nu)$ and the equality in the case of knots (\ref{pkknots}). 

 \begin{figure}[!ht]
\leavevmode
\begin{center}
\epsfysize=3cm
\epsfbox{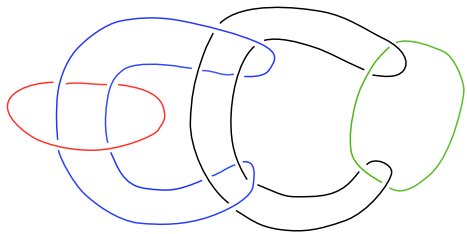}
\end{center}
\caption{A Brunnian link with four components.}
\label{brunnian}
\end{figure} 

The vanishing of $k_i^{(c),\CL}(\nu)$ with $i=0, \cdots, 3$ has been proved by Kanenobu in \cite{k}. More evidence for (\ref{genlinkvanish}) comes from Brunnian links. A Brunnian link is a nontrivial link with the property that every proper sublink is trivial. The Hopf link is a Brunnian link of two components, while the famous Borromean rings give a Brunnian link of three components. A Brunnian link with four components is 
shown in \figref{brunnian}. It is easy to see that the connected invariants of a Brunnian link $\CB$ with $L$ components are of the form 
\be\ba
\CG^{(c)}(\CB)&=\Bigl( 1+ {\nu -\nu^{-1} \over z} \Bigr) F_{\CB}(z,\nu)-\Bigl( 1+ {\nu -\nu^{-1} \over z} \Bigr)^L , \\
\CH^{(c)}(\CB)&={\nu -\nu^{-1} \over z} P_{\CB}(z,\nu)-\Bigl( {\nu -\nu^{-1} \over z} \Bigr)^L.
\ea
\ee
Conjectures (\ref{homflylink}) and (\ref{genlinkvanish}) imply that, for Brunnian links, 
\be
\label{brunnianeq}
P_{\CB}(z,\nu)-\Bigl( {\nu -\nu^{-1} \over z} \Bigr)^{L-1} =\CO(z^{L-1}), \qquad F_{\CB}(z,\nu)-\Bigl( 1+{\nu -\nu^{-1} \over z} \Bigr)^{L-1} =\CO(z^{L-1}).
\ee
This has been proved in \cite{pt,habiro}. 

 \begin{figure}[!ht]
\leavevmode
\begin{center}
\epsfysize=4cm
\epsfbox{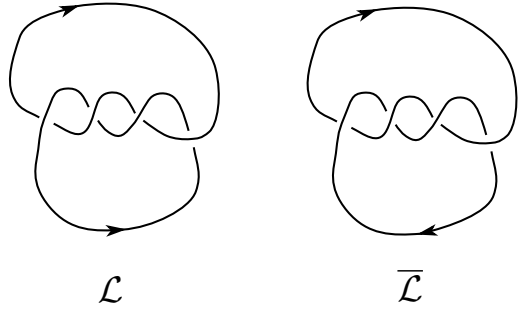}
\end{center}
\caption{The two oriented links $\CL$ and $\overline{\CL}$ tabulated as $4_1^2$, and differing in the relative orientation of their components.}
\label{pairl}
\end{figure} 

The relation (\ref{genlinkorientation}) (and in particular (\ref{ktwo}) for links with two components) 
seems however to be a new result in the theory of the Kauffman polynomial. It relates the Kauffman polynomial of an unoriented link $\CL$ 
to the HOMFLY polynomial of all the oriented links that can be obtained from $\CL$, modulo an overall reversal of the orientation. In the case of links made out of two unknots, it further simplifies to 
  \be
  \label{twous}
   k_2^{\CL}(\nu)=p_1^{\CL}(\nu) + p_1^{{\overline \CL}}(\nu)
   \ee
   and it can be easily checked in various cases by looking for example at the tables presented in \cite{lmtwo}. 
   
\begin{figure}[!ht]
\leavevmode
\begin{center}
\epsfysize=4cm
\epsfbox{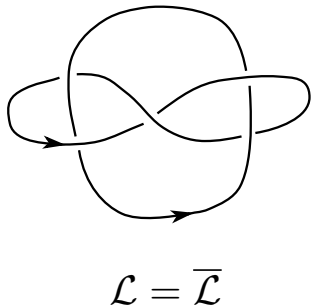}
\end{center}
\caption{The oriented link $\CL$, tabulated as $5_1^2$. It is invariant under reversal of orientation of its components, hence $\CL={\overline{\CL}}$.}
\label{singlel}
\end{figure}

   \begin{example} Let us check (\ref{twous}) for some simple links made out of two unknots. The easiest example is of course the Hopf link, where $\CL$ and ${\overline{\CL}}$ are 
   depicted in \figref{hreversed}. Their HOMFLY polynomials are given by 
   \be
   P_{\CL}= {\nu -\nu^{-1} \over z} +\nu z, \qquad    P_{\overline{\CL}}= {\nu -\nu^{-1} \over z} -\nu^{-1} z, 
   \ee
   and 
   \be
  p_1^{\CL}(\nu) + p_1^{{\overline \CL}}(\nu)=\nu-\nu^{-1}. 
  \ee
 By comparing with (\ref{khopf}), we see that (\ref{twous}) holds. Let us now consider the pair of oriented links depicted in \figref{pairl}. Their HOMFLY polynomials are 
 \be
 P_{\CL}=\bigl( \nu -\nu^{-1}\bigr) z^{-1} + \bigl( \nu -3 \nu^{-1}\bigr) z-\nu^{-1} z^3, \qquad  P_{{\overline{\CL}}}=\bigl( \nu -\nu^{-1}\bigr) z^{-1} + \bigl( \nu + \nu^3\bigr) z,
 \ee
 while the Kauffman polynomial is
 \be
 F_{\CL}= \bigl( \nu -\nu^{-1}\bigr) z^{-1} +1+ \bigl( \nu^3 +2 \nu -3 \nu^{-1}\bigr) z + \bigl( 1-\nu^2\bigr) z^{2} +\bigl( \nu -\nu^{-1}\bigr) z^3.
 \ee
 Again, the relation (\ref{twous}) holds. Finally, we consider the link depicted in \figref{singlel}, and tabulated as $5_1^2$. This link is invariant under reversal of orientation of its components, hence $\CL={\overline{\CL}}$, and its HOMFLY polynomial equals
 \be
 P_{\CL}=\bigl( \nu -\nu^{-1}\bigr) z^{-1} + \bigl(- \nu^{-1} +2 \nu - \nu^3 \bigr) z+\nu z^3,
 \ee
 while its Kauffman polynomial is 
 \be
 \ba
F_{\CL}&=\bigl( \nu -\nu^{-1}\bigr) z^{-1} +1+  \bigl(- 2 \nu^{-1} +4 \nu - 2 \nu^3 \bigr) z + \bigl( -1+\nu^4\bigr) z^{2} \\
&+\bigl( -\nu^{-1} + 3\nu -2\nu^3 \bigr) z^3 
+ (-1+ \nu^2)z^4.
\ea
\ee
Here, $k_2^{\CL}(\nu)=2 p_1^{\CL}(\nu)$, again in agreement with (\ref{twous}).  
\end{example}
 
Finally, it is easy to see that Rudolph's theorem for a link $\CL$ can be obtained by requiring integrality of, for example, the reformulated invariant $g_{\tableau{2}, \cdots, \tableau{2}}$, generalizing in this way our analysis for knots. It is likely that (\ref{mrgen}) follows from looking at $g_{S_1, \cdots, S_L}$ with $S_i \in R_i\otimes R_i$. 

\sectiono{String theory interpretation}

The conjecture stated in this paper is mostly based on the analysis performed in \cite{bfmtwo}, which in turn builds upon 
previous work on the large $N$ duality between Chern--Simons theory and topological strings (see \cite{mmreview} for a review of 
these developments). In this section we sketch some of the string theory considerations which lead to the above conjecture. For simplicity 
we will restrict ourselves to the case of knots. The extension of these considerations to the case of links is straightforward. 

\subsection{Chern--Simons theory and D-branes}

In \cite{wittenstring}, Witten showed that Chern--Simons theory on a three-manifold $M$ can be obtained by considering 
open topological strings on the cotangent space $T^*M$ with boundaries lying on $M$, which is a Lagrangian submanifold of $T^*M$. 
Equivalently, one can say that the 
theory describing $N$ topological branes wrapping $M$ inside $T^*M$ is ${\rm U}(N)$ Chern--Simons theory. 

To incorporate knots and links 
into this framework one has to introduce a different set of branes, as explained by Ooguri and Vafa \cite{ov}. This goes as follows: given any knot ${\cal K}$ in 
$\IS^3$, one can construct a natural Lagrangian
submanifold $\CN_{\cal K}$ in $T^*{\bf S}^3$. This construction is rather canonical, and it is called the conormal bundle
of ${\cal K}$. Let us parametrize the knot ${\cal K}$ by a curve $q(s)$, where $s\in [0, 2\pi]$. The
conormal bundle of ${\cal K}$ is the space
\be
\CN_{\cal K}= \Bigl\{ (q(s), p) \in T^*{\bf S}^3 \Bigl|
\sum_i p_i \dot q_i=0, \,\, 0 \le s \le  2\pi \Bigr\},
\end{equation}
where $q_i, p_i$ are coordinates for the base and the fibre of the cotangent bundle, respectively, and $\dot q_i$
denote derivatives w.r.t. $s$. The space $\CN_\CK$ is an $\IR^2$-fibration of the knot itself, where the fiber on the point $q(s)$
is given by the two-dimensional subspace of $T_{q(s)}^* \IS^3$ of planes
orthogonal to $\dot q(s)$. $ \CN_{\cal K}$ has the topology of $\IS^1 \times \IR^2$, and intersects $\IS^3$ along
the knot ${\cal K}$. As a matter of fact, for some aspects of the construction, 
the appropriate submanifolds to consider are deformations of $\CN_\CK$ which are disconnected from the zero section. 
For example, \cite{koshkin} considers a perturbation
\be
\label{epsdef}
\CN_{\CK, \epsilon}= \Bigl\{ (q(s), p+\epsilon \dot q(s)) \in T^*{\bf S}^3 \Bigl|
\sum_i p_i \dot q_i=0, \,\, 0 \le s\le 2\pi \Bigr\}.
\ee

Let us now wrap $M$ probe branes around $\CN_{\cal K}$. There will be open strings with one endpoint on $\IS^3$, and another endpoint on 
$\CN_{\cal K}$. These open strings lead to the insertion of the following operator (also called the Ooguri--Vafa operator) 
in the Chern--Simons theory on $\IS^3$ \cite{ov}:
\be
Z_{{\rm U}(N)} (v)=\sum_R  \tr_R^{{\rm U}(N)} (U_{\cal K}) \, s_R(v).
\ee
Here $U_{\cal K}$ is the holonomy of the Chern--Simons gauge field around ${\cal K}$, while $v$ is a $U(M)$ matrix associated to the $M$ branes wrapping 
$\CN_{\cal K}$. After computing the expectation value of this operator in Chern--Simons theory, we obtain the generating functional (\ref{zh}). 

In order to describe the Kauffman polynomial, we need a Chern--Simons theory on $\IS^3$ with gauge group ${\rm SO}(N)$ or ${\rm Sp}(N)$. From the point of 
view of the open string description, we need an {\it orientifold} of topological string theory on $T^*\IS^3$. This orientifold was constructed in \cite{sv}, and it can be 
described as follows. As a complex manifold, the cotangent space 
$T^*\IS^3$ is a Calabi--Yau manifold called the deformed conifold. It can be described by the equation 
\be
\sum_{i=1}^4 x_i^2=\mu,
\ee
where $x_i$ are complex coordinates. For real $\mu>0$, the submanifold ${\rm Im}\, x_i=0$ is nothing but $\IS^3$, while ${\rm Im}\, x_i$ are coordinates of the cotangent 
space. We now consider the following involution of the geometry
\be \label{stinv}
I: x_i \rightarrow \bar x_i
\ee
This leaves the $\IS^3$ invariant, and acts as a reflection on the coordinates of the fiber:
\be
p_i \rightarrow -p_i.
\ee
If we now wrap $N$ D-branes on $\IS^3$, the corresponding 
gauge theory description is 
Chern-Simons theory with gauge group ${\rm SO}(N)$ or ${\rm Sp}(N)$, depending on the choice of
orientifold action on the gauge group \cite{sv}. Since an orientifold theory is a particular case of a $\IZ_2$ orbifold, the partition function is expected to be 
the sum of the partition function in the untwisted sector, plus the partition function of the twisted sector. The partition function in the untwisted sector corresponds to a 
theory of oriented open strings in the ``covering geometry," i.e. the original target space geometry but with the closed moduli identified according to the 
action of the involution. The partition function in the twisted sector is given by the contributions of unoriented strings. We can then write
\cite{sv, bfmone}
\be
\label{covun}
Z_{{\rm SO}(N)/{\rm Sp}(N)}={1\over 2} Z_{\rm cov}+ Z_{\rm unor}.
\ee
In the case considered in \cite{sv}, the partition function for the covering geometry is just the ${\rm U}(N)$ partition function. 

 \begin{figure}[!ht]
\leavevmode
\begin{center}
\epsfysize=6cm
\epsfbox{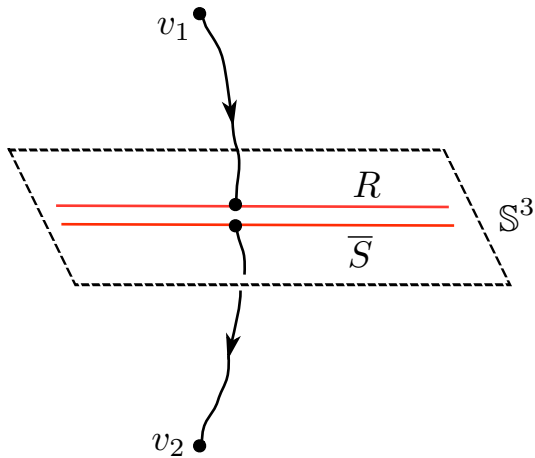}
\end{center}
\caption{The two sets of open strings in the covering geometry, going from the probe branes to the orientifold ``plane" in $\IS^3$ and extending along the cotangent directions. They are 
related by the target space involution, which sends $p_i \rightarrow -p_i$, and by orientation reversal. The Chan--Paton charges ending on $\IS^3$ lead to a Wilson 
line colored by $(R,\overline S)$, while the Chan-Paton charges in the probe branes lead to sources $v_1, v_2$ which have to be identified by the involution: $v_1=v_2$.}
\label{openstrings}
\end{figure} 

We can introduce Wilson loops around knots and links by following the strategy of Ooguri and Vafa, i.e. by introducing branes wrapping the Lagrangian submanifold $\CN_{\cal K}$. This leads to the insertion of the operator \cite{bfmone,bfmtwo}
\be
\label{soperator}
Z_{{\rm SO}(N)/{\rm Sp}(N)} (v)=\sum_R  \tr_R^{{\rm SO}(N)/{\rm Sp}(N)} (U_{\cal K}) \, s_R(v).
\ee
After computing the expectation value of this operator in Chern--Simons theory, we obtain the generating functional (\ref{zg}). In this paper we have used the Kauffman invariant which follows naturally from 
the gauge group ${\rm SO}(N)$, but in fact the two choices of gauge group lead to essentially identical theories 
due to the ``${\rm SO}(N) ={\rm Sp}(-N)$" equivalence, see \cite{bfmone} for a discussion and references. 

How does (\ref{soperator}) decompose into a sum (\ref{covun}) of covering and unoriented contributions? 
In general, if we have a geometry $X$ with a submanifold $L$, there will be {\it two} submanifolds in the covering geometry: the original submanifold $L$ and its image 
under the involution $I(L)$ \cite{bfmtwo}. Although $I( \CN_{\cal K}) =\CN_{\cal K}$, if one considers deformations of the conormal bundle, 
the resulting submanifolds will be different (this has been previously noticed in \cite{gw}). For example, for the deformation (\ref{epsdef}) one has 
that $I(\CN_{\CK,\epsilon})=\CN_{\CK,-\epsilon}$. Therefore, after deformation we will have {\it two} sets of probe D-branes in $T^*\IS^3$, wrapping two different submanifolds related by the involution $I$, 
and leading to two different sources $v_1$ and $v_2$ \cite{bfmtwo}. In particular, we have {\it two} sets of open strings, going from the two sets of probe branes to the branes wrapping $\IS^3$ in the orientifold plane, and related by the orientifold action. This action involves both the target space involution $p_i \rightarrow -p_i$ and an orientation reversal which conjugates the Chan-Paton charges. We then conclude that one set of open strings will lead to the insertion of Wilson lines in $\IS^3$ involving representations $R=\cdot, \tableau{1}, \tableau{2}, \cdots$, while the other set of open strings will lead to {\it conjugate} representations $S=\cdot, {\overline {\tableau{1}}}, {\overline{\tableau{2}}}, \cdots$. This is illustrated in \figref{openstrings}. 

The partition function of the covering geometry will then have the structure
\be
\label{zcovv}
Z_{\rm cov}(v)=\sum_{R, S} \tr^{{\rm U}(N)}_{(R, S)}(U_{\CK})\, s_{R}(v) s_{S}(v).
\ee
Since we have to identify the closed and open moduli according to the action of the involution, in (\ref{zcovv}) 
we have set $v_1=v_2=v$ in the source terms $s_{R}(v_1)$ and 
$s_{S}(v_2)$. After computing the expectation value of this operator in 
Chern--Simons theory, we obtain the generating functional $Z_\CR$ in (\ref{zr}). 

This argument by itself does not make possible to decide if the representation induced 
by the orientifold action is the composite representation $(R, S)$ or the tensor product representation $R\otimes {\overline {S}}$, which differ in ``lower order corrections" 
as specified in (\ref{tensorcomposite}). One needs in principle a more detailed study of the orientifold, but as we will see in a moment, by looking at the topological string theory 
realization for simple knots and links, we can verify that the covering geometry involves indeed the composite representation.

\subsection{Topological string dual}

It was conjectured in \cite{gv} that open topological string theory on $T^*\IS^3$ with $N$ D-branes wrapping $\IS^3$ is equivalent to a closed topological string theory on the resolved 
conifold
\be
\label{rescon}
X=\CO(-1) \oplus \CO(-1) \rightarrow \IP^1. 
\ee
This leads to a large $N$ duality between Chern--Simons theory on $\IS^3$ and the closed topological string on (\ref{rescon}). The open string theory on the deformed conifold is 
related to the closed string theory on the resolved conifold by a so-called {\it geometric transition}. In this case this is the conifold transition. 

This duality was extended by Ooguri and Vafa 
to the situation in which one has probe branes in $T^*\IS^3$ wrapping the Lagrangian submanifold $\CN_{\CK}$. They postulated that, given any knot $\CK$, one 
can construct a Lagrangian submanifold in the resolved conifold, $\CL_K$, which can be understood as a geometric transition of the Lagrangian $\CN_{\CK}$ in the 
deformed conifold. The total free energy in the deformed conifold can be computed in terms of Chern--Simons theory and it is given by $F_{\CH}(v)$. By the large $N$ duality, it should be equal to
 the free energy of an open topological string theory on the resolved conifold $X$ with Lagrangian boundary conditions given by $\CL_K$. 
 Since open topological string amplitudes can be reformulated in terms of counting of BPS invariants and 
satisfy integrality conditions \cite{gvonetwo,ov,lmv}, one obtains the conjecture about the integrality properties of the colored HOMFLY invariant \cite{lmknot,mmreview} which we reviewed above. 

As first shown in \cite{sv}, one can extend the large $N$ duality of \cite{gv} to the orientifold case and obtain an equivalence between Chern--Simons theory on  $\IS^3$ with ${\rm SO}(N)/{\rm Sp}(N)$ gauge groups and topological string theory on an orientifold of the resolved conifold. A convenient description of (\ref{rescon}) is as a toric manfiold, defined by the 
equation
\be
|X_1|^2 + |X_2|^2 -|X_3|^2 - |X_4|^2 =t
\ee
and a further quotient by a $U(1)$ action where the coordinates $(X_1, \cdots, X_4)$ have charges $(1,1,-1,-1)$. In this description, the orientifold is defined by the 
involution
\be
\label{orres}
(X_1, X_2, X_3, X_4) \rightarrow ({\overline X}_2, -{\overline X}_1,
{\overline X}_4, -{\overline X}_3).
\ee

Let us now consider the open topological string theory on the resolved conifold defined by the Lagrangian submanifold $\CL_K$ associated to a knot. If we perform the orientifold action (\ref{orres}) we obtain an orientifold of this open topological string theory. The total partition function of this orientifold of the resolved conifold should be equal to the total partition function of the orientifold of the deformed conifold, 
namely $Z_\CG (v)$. The contribution of the covering geometry will be given by the partition function of 
topological open strings on $X$ in the presence of two Lagrangian submanifolds, $\CL_K$ and $I(\CL_K)$, after identifying the sources. This should be equal to the contribution of the 
covering of the deformed geometry, i.e. $Z_{\CR}(v)$. This partition function can then be expressed in terms of integer BPS invariants $N_{R;g,Q}^{c=0}$. 

On the other hand, the unoriented contribution to the orientifold partition function 
will be given by the partition function of unoriented topological open strings 
in the quotient geometry $X/I$ with a brane $\CL_K$. This unoriented partition function also has an integrality structure \cite{bfmone,bfmtwo} generalizing \cite{gvonetwo}. In particular, it can be written in terms of BPS invariants $N_{R;g,Q}^{c=1,2}$ related to the counting of curves with boundaries ending on $\CL_K$ and with one or two crosscaps. This 
explains the integrality properties for the colored Kauffman polynomial that we conjectured in this paper. 

\begin{remark} Note that the two choices of orientifold action which lead to the gauge groups 
${\rm SO}(N)/{\rm Sp}(N)$ in the deformed conifold become here a choice of overall sign for the $c=1$ contribution, see for example \cite{bfmone} for a discussion and examples. 
\end{remark}
\begin{remark} The sum over odd positive integers $d$ in (\ref{grdef}) seems to be a general feature of multicovering formulae for unoriented surfaces, as noticed in \cite{sv,bfmone,bfmtwo}. See 
\cite{kw} for recent examples. 
\end{remark}

When $\CK$ is the unknot it is possible to construct explicitly the corresponding Lagrangian submanifold $\CL_K$ in $X$. It turns out to be given by a toric construction, and it is possible to compute 
$Z_{\CR}(v)$ by using the topological vertex \cite{bfmtwo}. The explicit computation in equation (3.10) of \cite{bfmtwo} 
confirms that the vacuum expectation value of the operator appearing in (\ref{zcovv}) is indeed the 
quantum dimension of the composite representation (\ref{qdimcr}),
\be
\langle   \tr^{{\rm U}(N)}_{(R, S)}(U_{\CK}) \rangle ={\rm dim}_q \, (R,S).
\ee
One can also find an explicit description of the Lagrangian submanifolds in $X$ corresponding to the Hopf link \cite{bfmtwo}, and 
compute the covering contribution to the orientifold 
partition function by using the 
topological vertex. The resulting expression (equation (3.17) of \cite{bfmtwo}) agrees again with the HOMFLY invariant of the Hopf link for general composite representations, which was computed in \cite{anv,kanno} in a different context. 

\sectiono{Conclusions and outlook}

In this paper we have formulated a new conjecture on the structure of the colored Kauffman polynomial of knots and links. This conjecture is mainly based on the results of \cite{bfmtwo}, but it adds a crucial ingredient which was missing in that paper: the fact that partition functions in the untwisted sector of the orientifold are given by HOMFLY invariants colored by composite representations. This makes possible to extend the results obtained for the colored HOMFLY invariant in \cite{ov,lmv,lmknot} to the colored Kauffman polynomial. According to our conjecture, the natural invariant of unoriented knots 
and links involves both the Kauffman polynomial and the HOMFLY polynomial colored with composite representations. In particular, 
in the case of links, it involves considering all possible orientations for the 
components of a link. This is probably the most interesting aspect of the conjecture, and it ``explains" many aspects of the relationship between these invariants, like Rudolph's theorem \cite{rudolph}. It also leads to new, simple results about the Kauffman polynomial, like for example (\ref{genlinkorientation}). From the point of 
view of physics, the results presented in this paper provide new precision tests of a large $N$ string/gauge theory correspondence.

It would be very interesting to 
relate the integrality properties conjectured here to appropriate generalizations of Khovanov homology, as in \cite{gsv}. Indeed, as in the case of the colored HOMFLY invariants, 
the integers $N_{R_1, \cdots, R_L;g,Q}^{c=0,1,2}$ are Euler characteristics of cohomology theories associated to 
BPS states, and it is natural to conjecture that these cohomologies give categorifications of the colored Kauffman invariant. There has been 
already work in this direction for knots colored by the fundamental representation \cite{gw}. The case of links and/or higher representations should involve, 
as conjectured in this paper, both the Kauffman invariant and the HOMFLY invariant for composite 
representations. 

Finally, it was noticed in \cite{lm} that  the reformulated invariants $f_R$, when expanded in power series, $t=\exp(x/2)$, lead to Vassiliev invariants. On the other hand, some of the properties that follow from our conjectures (like 
(\ref{brunnianeq})) have a natural interpretation in Vassiliev theory. Therefore, it would be interesting to have a 
precise interpretation of our conjectures in terms of Vassiliev 
invariants, specially now that we have a rather complete understanding of Chern--Simons invariants for all classical gauge groups in terms of string theory.  

\vskip .5cm

{\bf Note added}: After this paper was submitted, two papers appeared \cite{stevan,ramacom} with extensive checks of conjecture \ref{myconj} for 
framed torus knots and links. 
 
\section*{Acknowledgments}
My interest in this problem was revived by the recent paper of Morton and Ryder \cite{mr}, and I would like to thank them for a very useful correspondence. I would also like to thank Vincent Bouchard, 
Jose Labastida and Cumrun Vafa for many conversations on this topic along the years, and S\'ebastien Stevan for recent discussions on torus knots. Finally, I would like to thank Vincent Bouchard, Stavros Garoufalidis and specially Hugh Morton for a detailed reading of the manuscript. This work was supported in part by the Fonds National Suisse.

\end{document}